\documentstyle[12pt,psfig]{article}

\newcommand{\ba}{\begin{eqnarray}}
\newcommand{\ea}{\end{eqnarray}}
\newcommand{\half}{{\textstyle{\frac{1}{2}}}}
\newcommand{\partialslash}{\partial\hspace{-.5em}/\hspace{.15em}}
\newcommand{\Pslash}{P\hspace{-.7em}/\hspace{.25em}}
\newcommand{\Qslash}{Q\hspace{-.7em}/\hspace{.25em}}

\newcommand{\kslash}{k\hspace{-.5em}/\hspace{.15em}}
\newcommand{\bfk}{{\bf k}}
\newcommand{\kint}{\int_{\Lambda_3}\!\frac{d^4 k}{(2\pi)^4}}

\newsavebox{\s}
\newsavebox{\tmpbox}
\newcommand{\Slash}[1]{\sbox{\s}{#1} \hbox to \wd\s {#1\hss\hbox to \wd\s{\hss/\hss}}}
\newcommand{\fourint}[1]{\int\!\frac{d^4 #1}{(2\pi)^4}}

\newcommand{\be}{\begin{equation}}
\newcommand{\ee}{\end{equation}}

\newcommand{\kstar}{K_{0}^{*}}
\newcommand{\diag}{{\rm diag}}
\newcommand{\J}{{\cal J}}

\newcommand{\Tr}{{\rm Tr}}
\newcommand{\mev}{{\kern2mm\rm MeV}}
\newcommand{\gev}{{\kern2mm\rm GeV}}

\newcommand{\journal}[5]{{\bf #1}---  #2, #4,  v.~#3, p.~#5}

\begin{document}

\begin{titlepage}
\title{\Large\bf Radially excited scalar, pseudoscalar,
and vector meson nonets in a chiral quark model. } 
\author{\large\bf M.K.\ Volkov, V.L.Yudichev \\[0.4cm]
\em Bogoliubov Laboratory of Theoretical Physics \\
\em Joint Institute for Nuclear Research \\
\em 141980, Dubna, Moscow region, Russia}
\date{}
\end{titlepage}
\maketitle
\begin{abstract}
A chiral Lagrangian containing, besides the usual meson
fields, their first radial excitations  is constructed. The
Lagrangian is derived by bosonization of a Nambu--Jona-Lasinio (NJL) type  quark
model with separable nonlocal interactions.
The nonlocality is described by form factors
corresponding to 3-dimensional  excited state wave
functions. The spontaneous breaking of chiral symmetry
is governed by the standard NJL gap equations.
A simple $SU(2)\times SU(2)$ version of the model is used to
demonstrate    all low-energy theorems to hold valid in the
chiral limit.

A more realistic $U(3)\times U(3)$ model
with 't Hooft interaction is
constructed to describe the mass spectrum of excited scalar,
pseudoscalar, and vector mesons. On the basis of
global chiral symmetry, we use the same form factors for
the scalar and pseudoscalar mesons. Having fixed the form factor
parameters by  masses of pseudoscalar
mesons, we predict the mass spectrum of scalar mesons. This
allows us to interpret experimentally observed scalar, pseudoscalar, and
vector  meson states
as members of quark-antiquark nonets. It is shown that
the $a_0(1450)$, $K^*_0(1430)$, $f_0(1370)$, $f_J(1710)$
scalar meson states are the first radial excitations of
the ground states: $a_0(980)$, $K^*_0(960)$, $f_0(400-1200)$,
$f_0(980)$.
The weak decay constants $F_\pi$,
$F_{\pi'}$, $F_K$, $F_{K'}$ and the main strong decay widths of
the scalar, pseudoscalar, and vector meson nonets are calculated.
\end{abstract}

\newpage

\section{Introduction}

The investigation of radial excitations of the scalar, pseudoscalar,
and vector  meson nonets is of great
interest in the hadronic physics.
So far, there are  questions connected
with the experimental and theoretical descriptions of 
radial excitations of scalar and pseudoscalar mesons.
For instance, the experimental data on the
excited states of kaons \cite{PDG} are rare and
not  reliable enough.
There are also  problems with
interpretation of the experimental data on the scalar and
$\eta$, $\eta'$ mesons. Several years ago,
attempts were undertaken to consider the state $\eta'(1440)$ as a
glueball \cite{geras}.

There is an analogous problem
with  interpretation of the scalar states $f_0(1500)$ and $f_0(1710)$.
Moreover, the experimental
     status of the lightest scalar isoscalar  singlet meson
     remains unclear. In some papers, the resonance
     $f_0(1370)$ was considered as a member of the ground nonet
     \cite{F1370,dmitr}, and  until 1998  the resonance
     $f_0(400-1200)$ was not included into the
     summary tables of PDG review%
     \footnote{ However, in earlier editions
     of PDG  the light $\sigma$ state could still be
     found;  it was excluded later.  }
     \cite{PDG}.

One will find a problem of the same sort in the case of $\kstar$.
The strange meson
     $K^*_0(1430)$ seems too heavy to be the ground state: $1\gev$
     is more characteristic of the ground meson states (see \cite{ishida,scadron}).

Anticipating the results of our review we would like to note 
that some of these problems were solved in a number of our
works which resulted in the present work.
From our calculations, for example, we concluded 
that the states $\eta(1295)$
and $\eta(1440)$ can be considered as radial excitations of the
ground states $\eta$ and $\eta'$.
The estimates of their strong decay widths also confirm
 our conclusion. Let us note that these meson states are
essentially mixed.
Our calculations  also showed
that we can interpret the scalar states $f_0(1370)$, $a_0(1450)$, $f_0(1710)$,
and $\kstar(1430)$
as the first radial excitations of $f_0(400-1200)$, $a_0(980)$, $f_0(980)$, and
$\kstar(960)$.

A theoretical description of radially excited pions poses some
interesting challenges. The physics of normal pions is completely
governed by the spontaneous breaking of chiral symmetry (SBCS). A convenient
way to derive the properties of soft pions is the use of an effective
Lagrangian based on a non-linear or linear realization of chiral symmetry
\cite{CCWZ69}. When attempting to introduce higher resonances to
extend the effective Lagrangian description to higher energies, one
must ensure that the introduction of new degrees of freedom does not
spoil the low--energy theorems for pions which are universal
consequences of chiral symmetry.

Attempts to describe heavier analogs of the pion, vector mesons,
and $\eta$, $\eta'$ mesons as the radial excitations of
 well-known ground meson states were made by authors in
\cite{geras} within the framework of the nonlocal
${}^3P_0$ potential quark model. This approach was based
on non-relativistic and relativistic quantum mechanics where
mesons are treated as bound $\bar qq$ systems.

A useful guideline in the construction of effective meson Lagrangians
is the Nambu--Jona-Lasinio (NJL) model that describes 
SBCS at the quark level with a
four--fermion interaction
\cite{volkov_83,volk_84,volk_86,ebert_86}.
The bosonization of this model and the derivative expansion
of the resulting fermion
determinant reproduce the Lagrangian of the linear sigma model that
embodies the physics of soft pions as well as higher--derivative
terms.  With appropriate couplings the model allows one to derive also a
Lagrangian for vector and axial--vector mesons.
This  gives not only the correct
structure of terms of the Lagrangian as required by chiral
symmetry, but also quantitative predictions for the
coefficients, such as $F_\pi$, $F_K$, $g_\pi$, $g_\rho$, {\em etc}.

One may, therefore, hope that a
suitable generalization of the NJL--model can provide  means for
deriving an effective Lagrangian including also the excited mesons.

When extending the NJL model to describe radial excitations of
mesons, one has to introduce nonlocal (finite--range) four--fermion
interactions.  Many nonlocal generalizations of the NJL model were
proposed, by using either covariant--Euclidean \cite{roberts_88} or
instantaneous (potential--type) \cite{leyaouanc_84,pervushin_90}
effective quark interactions.  These models generally require bilocal
meson fields for bosonization, which makes it difficult to perform a
consistent derivative expansion leading to an effective
Lagrangian.

A simple alternative is to use  separable
quark interactions. There is a number of advantages of working with
that scheme. First, separable interactions can be bosonized by
introducing local meson fields, just as the usual
NJL--model. One can thus derive an effective meson
Lagrangian directly in terms of local
fields and their derivatives. Second, separable interactions allow
one to introduce a limited number of excited states and only in a
given channel.

An interesting method for describing excited meson
states in this approximation was proposed in \cite{andrianov_93}.
The authors suggested to consider SBCS
in the vicinity of a polycritical point where either all or some of
the coupling constants at four-fermion vertices exhibit critical
behavior; the critical values of the coupling constants are given
by  solutions of a set of mass-gap equations.
They  selected a minimal type of separable
four-quark interaction which is most important for the
process of SBCS.
In this model the  form factors are chosen
as orthogonal functions, so there is a freedom in their choice
up to an arbitrary rotation.
All calculations are made in the Euclidean space, by using the
 approximation of large $N_c$ and
$\log\Lambda$ where $\Lambda$ is the ultra-violet cut-off in
the model. An interesting result of
this approach is that for an arbitrary choice of coupling
constants in the vicinity of polycritical point there are multiple
solutions with a different critical behavior. Therefore,
a problem appears --- which of the solutions is realized in
nature.

Another advantage of the separable interaction is that it 
can be defined in Minkowski
space in a 3--dimensional (yet covariant) way, with form factors
depending only on part of the quark--antiquark relative momentum
transverse to the meson momentum
\cite{pervushin_90,kalinovsky_89,weiss}.
This is essential for a correct description of excited states, since
it ensures the absence of spurious relative--time excitations
\cite{feynman_71}.  Finally, as we have shown \cite{weiss},
the form factors
defining the separable interaction can be chosen so that
the gap equation of the generalized NJL--model coincides with the one
of the usual NJL--model, whose solution is a constant
(momentum--independent) dynamic quark mass. Thus, in this approach
it is possible to describe radially excited mesons above the usual
NJL vacuum. Aside from the technical simplification, the latter means
that the separable generalization contains all the successful
quantitative results of the usual NJL model.

Our work consists of five Sections. In the second Section,
we illustrate our method on the basis of a simple
$SU(2)\times SU(2)$ model. Here we prepare grounds for the
choice of the form factors to be used in a more realistic model.
It will be shown  that we can choose these form factors such that
the gap equation conserves its conventional form and has a solution
corresponding to a constant constituent quark mass.
The quark condensate also does not change  after the inclusion
of excited states into the model, because the tadpole associated
with the excited scalar field is equal to zero (the quark loop
with the one excited scalar vertex, vertex with  a form factor).

In this Section, we derive an effective chiral Lagrangian describing
$\pi$ and $\pi'$ mesons from a generalized NJL--model with separable
interactions. In Subsection 2.1, we introduce the effective quark
interaction in the separable approximation and describe its
bosonization.  We discuss the choice of form factors necessary to
describe excited states. In Subsection 2.2, we solve the gap equation
defining the vacuum, derive the effective Lagrangian of the $0^-$
meson fields, and perform the diagonalization leading to the physical
$\pi$ and $\pi'$ states. The effective Lagrangian describes the
vanishing of the $\pi$ mass (decoupling of the Goldstone boson) in
the chiral limit, while $\pi'$ remains massive.  In Subsection 2.3, we
derive the axial vector current of the effective Lagrangian using the
Gell-Mann--Levy method and obtain a generalization of the PCAC
formula which includes the contribution of $\pi'$ to the axial
current. The leptonic decay constants of the $\pi$ and $\pi'$ mesons,
$F_\pi$ and $F_{\pi'}$, are discussed in Subsection 2.4.  It is shown that
$F_{\pi'}$ vanishes in the chiral limit as expected.  In Subsection 2.5,
we fix the parameters of the model and evaluate the ratio
$F_{\pi'} / F_\pi$ as a function of the $\pi'$ mass.

In the third Section, we use the method demonstrated in Section 2
for a  realistic description  of radially excited states of
the scalar, pseudoscalar and vector meson nonets where 
't Hooft interaction is included in addition
to conventional chirally symmetric four-quark vertices.
This  allows us to solve the so-called $U_A(1)$-problem and
describe the masses of ground and excited states of
the $\eta$ and $\eta'$ mesons .

 We take 
account of the connections of the scalar and vector coupling
constants which  appeared in this model and an additional
renormalization of the pseudoscalar fields connected with
the pseudoscalar --- axial-vector transitions.
For simplicity, we  suppose that the masses of $u$
and $d$ quarks are equal to each other and take into
account only the mass difference between ($u$, $d$)
and $s$ quarks ($m_u$ and $m_s$). Then, we have
in this model  six basic parameters: $m_u$, $m_s$,
$\Lambda_3$ (3-dimensional cut-off parameter), $G$
and $G_V$ (four--quark coupling constants for
the scalar--pseudoscalar coupling ($G$) and for the
vector -- axial--vector coupling ($G_V$)) and constant $K$
characterizing the 't Hooft interaction. To  define
these parameters, we use the experimental values: the
pion decay constant $F_\pi = 93 \mev$, the $\rho$--meson decay
constant $g_\rho \approx 6.14$ ($g_\rho^2/(4\pi)
\approx 3$), the pion mass $M_\pi \approx 140 \mev$,
$\rho$-meson mass $M_\rho = 770 \mev$, the kaon mass
$M_K \approx 495 \mev$, and the mass difference of the
$\eta$ and $\eta'$ mesons. Using these six parameters, we can
describe the masses of four ground meson nonets (pseudoscalar,
vector, scalar, and axial--vector)
and all the meson coupling
constants of strong interactions of
mesons with each other and with  quarks.

For the investigation of  excited states of the mesons it
is necessary to consider  nonlocal four--quark interactions.
In Section 3, it is shown that for the
description of excited states of the scalar, pseudoscalar,
 and vector meson nonets we have to use
seven different form factors in the effective
four-quark interactions.
Each form factor contains only one
free (external) parameter. There are also slope (internal)
parameters which are to be fixed by the condition
of preserving gap equations in the standard form (see Section 2).
We use the same form factors for the scalar and pseudoscalar mesons,
which is required by chiral symmetry. This 
allows us to predict masses of the excited scalar mesons.

In Subsection 3.1, we introduce the effective quark interaction
in the separable approximation with 't Hooft terms
and describe its bosonization.
We discuss the choice of the form factors
necessary to describe  excited states of the scalar,
pseudoscalar, and vector meson nonets. In Subsection 3.2, we derive
the effective Lagrangian for the ground and excited states of the 
strange and isovector scalar and pseudoscalar mesons, and
perform the diagonalization leading to the physical 
ground and excited meson states.
In Subsection 3.3, we diagonalize the Lagrangian for the 
isoscalar scalar and pseudoscalar (ground and excited) 
mesons and take into account  singlet-octet mixing.
In Subsection 3.4,
we consider vector mesons. In Subsection 3.5, we fix the
parameters of the model and evaluate the masses of the ground and
excited meson states and the weak decay constants $F_{\pi}$, $F_{\pi'}$,
$F_K$ and $F_{K'}$.

In Section 4, we calculate strong decay widths of  excited
states of the scalar, pseudoscalar, and vector mesons and compare them
with experimental data.
In Subsection 4.1, we consider decays of the first radial excitations of
$\pi$, $\rho$ and $\omega$ meson states. Decays of strange mesons
are calculated in Subsection 4.2.
Then, in Subsection 4.3, we calculate
decay widths of the scalar mesons.
Finally, the decay widths of excited $\eta$ and $\eta'$ mesons
are estimated in Subsection 4.4.

In Section 5 (Conclusion), we  briefly discuss our results,
give interpretation of the members of meson nonets, and
foresee ways of further developing our model.

In Appendix A, we collected some lengthy formulae defining
the free part of the effective Lagrangian for isoscalar 
scalar and pseudoscalar mesons. In Appendix B, we displayed in detail
some instructive calculations of  strong decay widths of mesons.


\section{$SU(2)\times SU(2)$ model.}

\subsection{Nambu--Jona-Lasinio model with separable interactions}
In this Section, we construct an $SU(2)\times SU(2)$ NJL-like chiral 
quark model with quark interaction of the separable type to describe
the ground and first radially excited states of pions and $\sigma$-mesons.
 Although,  a realistic description
of the meson physics requires consideration of  a $U(3)\times U(3)$ version
(which we will do in the next Section), we find it instructive to show
the basic principles of the model with this simple case. The content of
the section corresponds to ref.~\cite{weiss}.

In the usual NJL model, SBCS
is described by a local (current--current) effective quark
interaction. The model is defined by the action
\ba
S [\bar\psi , \psi ] &=& 
\int d^4 x \, \bar\psi (x) \left( i \partialslash - m^0 \right)
\psi (x) \; + \; S_{\rm int} ,\label{S_NJL} \\
S_{\rm int} &=& \frac{G}{2} \int d^4 x \left[ j_\sigma (x) j_\sigma (x) + 
j_\pi^a (x) j_\pi^a (x) \right],
\label{S_int}
\ea
where $j_{\sigma , \pi} (x)$ denote, respectively, the 
scalar--isoscalar and pseudoscalar--isovector currents of the 
quark fields ($SU(2)$--flavor),
\be
j_\sigma (x)= \bar\psi (x) \psi (x), \hspace{2cm}
j_\pi^a (x) \;\; = \;\; \bar\psi (x) i\gamma_5 \lambda^a \psi (x) .
\label{j_def}
\ee
The model can be bosonized in a standard way by representing the
4--fermion interaction as a Gaussian functional integral over scalar
and pseudoscalar meson fields \cite{volkov_83,volk_84,volk_86,ebert_86}. Since the
interaction, Eq.(\ref{S_int}), has represents a product of two local
currents, the bosonization is achieved through local meson
fields. The effective meson action  obtained by integration
over quark fields is thus expressed in terms of local meson
fields. By expanding the quark determinant in derivatives of the
local meson fields, one then derives the chiral meson Lagrangian.
\par
The NJL interaction, Eq.(\ref{S_int}), describes only ground--state
mesons. To include excited states, one has to introduce effective
quark interactions with a finite range.  In general, such
interactions require bilocal meson fields for bosonization
\cite{roberts_88,pervushin_90}. A possibility to avoid this
complication is ot use a separable interaction that  is still of
current--current form, Eq.(\ref{S_int}), but allows for nonlocal
vertices (form factors) in the definition of  quark currents,
Eq.(\ref{j_def}),
\ba
\tilde{S}_{\rm int} &=& \frac{G}{2} \int d^4 x \sum_{i = 1}^N
\left[ j_{\sigma , i} (x) j_{\sigma , i} (x) 
+ j_{\pi , i}^{a} (x) j_{\pi , i}^{a} (x) \right] , 
\label{int_sep}
\\
j_{\sigma , i} (x) &=& \int d^4 x_1 \int d^4 x_2 \; 
\bar\psi (x_1 ) F_{\sigma , i} (x; x_1, x_2 ) \psi (x_2 ), 
\label{j_sigma} \\
j_{\pi , i}^{a} (x) &=& \int d^4 x_1 \int d^4 x_2 \; 
\bar\psi (x_1 ) F_{\pi , i}^{a} (x; x_1, x_2 ) \psi (x_2 ) . 
\label{j_pi} 
\ea
Here, $F_{\sigma , i}(x; x_1, x_2 ), F_{\pi , i}^{a}(x; x_1, x_2 ), 
\, i = 1, \ldots N$, denote a set of nonlocal scalar and
pseudoscalar fermion vertices (in general, momentum-- and
spin--dependent) to be specified below. Upon bosonization
Eq.(\ref{int_sep}) leads to the action
\ba
&&S_{\rm bos}[\bar\psi , \psi; \sigma_1 , \pi_1 , \ldots 
\sigma_N , \pi_N ]  
= \int d^4 x_1 \int d^4 x_2 \;
\bar\psi (x_1 ) \left[ \left( i \partialslash_{x_2} - m^0 \right) 
\delta (x_1 - x_2 ) \rule{0cm}{1.5em}
\right. \nonumber \\
&&\quad \left. \rule{0cm}{1.5em}
+ \int d^4 x  \sum_{i = 1}^N \left( \sigma_i (x)
F_{\sigma , i} (x; x_1, x_2 ) + \pi_i^a (x) 
F_{\pi , i}^a (x; x_1, x_2 )
\right) \right] \psi (x_2 ) \nonumber \\
 &&\quad- \frac{1}{2G} \int\! d^4 x \sum_{i = 1}^{N}
\left( \sigma_i^2 (x) + \pi_i^{a\, 2} (x) \right) .
\label{S_sep}
\ea
It describes a system of local meson fields, 
$\sigma_i (x), \pi_i^a (x),\, i = 1, \ldots N$, which interact with
quarks through nonlocal vertices. We emphasize that these fields
are not yet to be associated with physical particles ($\sigma ,
\sigma', \ldots, \pi, \pi' , \ldots$); physical fields will be
obtained after determining the vacuum and diagonalizing the 
meson effective action.

To define the vertices of Eqs.(\ref{j_sigma}) and (\ref{j_pi}), and we pass to
the momentum representation. Because of the translational invariance, the
vertices can be represented as
\ba
\lefteqn{ F_{\sigma , i} (x; x_1, x_2 ) } && \nonumber \\
&=& \fourint{P} \fourint{k}
\exp i \left[ \frac12 (P + k) \cdot (x - x_1 )
+ \frac12 (P - k) \cdot (x - x_2 ) \right]
F_{\sigma , i} (k | P) , \nonumber
\\
\ea
and similarly for $F_{\pi , i}^{a} (x; x_1, x_2 )$. Here $k$ and $P$
denote, respectively, the relative and total momentum of a
quark--antiquark pair. We take the vertices to depend only on the
component of the relative momentum transverse to the total momentum,
\be
F_{\sigma , i} (k | P) \equiv
F_{\sigma , i} (k_\perp | P), \hspace{1cm} \mbox{\it etc.}, 
\hspace{2cm}
k_\perp \; \equiv \; k - \frac{P\cdot k}{P^2} P.
\label{markov_yukawa}
\ee
Here, $P$ is assumed to be time-like, $P^2 > 0$.
Equation(\ref{markov_yukawa}) is a covariant generalization of the
condition that the quark--meson interaction is instantaneuos in the
rest frame of the meson ({\em i.e.}, the frame in which 
${\bf P} = 0$). Equation (\ref{markov_yukawa}) ensures the absence of
spurious relative--time excitations and thus leads to a consistent
description of excited states\footnote{In bilocal field theory, this
requirement is usually imposed in the form of the so--called
Markov--Yukawa condition of covariant instanteneity of the bound
state amplitude \cite{pervushin_90}. An interaction of the transverse
form, Eq.(\ref{markov_yukawa}), automatically leads to meson
amplitudes satisfying the Markov--Yukawa condition.}
\cite{feynman_71}.  In particular, this framework allows us to use
3--dimensional ``excited state'' wave functions to model the form
factors for radially excited mesons.

The simplest chirally invariant interaction describing scalar and
pseudoscalar mesons is defined by  spin--independent vertices $1$
and $i\gamma_5 \lambda^a$, respectively. We want to include ground state
mesons and their first radial excitation ($N = 2$), and therefore
take
\ba
\left.
\begin{array}{r}
F_{\sigma , j} (k_\perp | P)  \\ 
F_{\pi , j}^{a} (k_\perp | P) 
\end{array} \right\} 
&=& 
\left\{
\begin{array}{r}
1 \\              
i \gamma_5 \lambda^a 
\end{array} \right\}
\times \Theta (\Lambda_3 - | k_\perp | ) \, f_j(k_\perp )
\label{F_2}, 
\\
f_1(k_\perp )\equiv 1,\qquad
f_2(k_\perp ) &=& c (1 + d \, | k_\perp |^2 ) , 
\qquad | k_\perp | \; \equiv \; \sqrt{-k_\perp^2}.
\label{f}
\ea
The step function, $\Theta (\Lambda_3 - | k_\perp | )$, is nothing
else then a covariant generalization of the usual 3--momentum cutoff of
the NJL model in the meson rest frame \cite{pervushin_90}. The form
factor $f(k_\perp )$ has for $d < -\Lambda_3^{-2}$ the form of an
excited state wave function, with a node in the interval 
$0 < | k_\perp | < \Lambda_3$.  Equations (\ref{F_2}) and (\ref{f})
are the first two terms in a series of polynomials in $k_\perp^2$;
inclusion of higher excited states would require polynomials of
higher degree.  Note that the normalization of the form factor
$f(k_\perp )$, the constant $c$, determines the overall strength of
the coupling of the $\sigma_2$ and $\pi_2$ fields to  quarks
relative to the usual NJL--coupling of $\pi_1$ and $\sigma_1$.

We remark that the most general vertex could also include
spin--dependent structures, $\Pslash$ and $\gamma_5\Pslash$, which in
the terminology of the NJL model correspond to the induced vector and
axial vector component of $\sigma$ and $\pi$ ($\sigma$--$\rho$
and $\pi$--$A_1$ mixing), respectively. These structures should be
considered if vector mesons are included. Furthermore, there could be
structures $\kslash_\perp , \Pslash \kslash_\perp$ and
$\gamma_5\kslash_\perp , \gamma_5\Pslash \kslash_\perp$,
respectively, which describe bound states with orbital angular
momentum $L = 1$.  We shall not consider these components here.

With the form factors defined by Eqs.(\ref{F_2}) and (\ref{f}),
the bosonized action, Eq.(\ref{S_sep}), in the momentum representation
takes the form
\ba
&&S_{\rm bos}[\bar\psi , \psi; \sigma_1 , \pi_1 , 
\sigma_2 , \pi_2 ]   
= \fourint{k} \bar\psi (k) \left( \kslash - m^0 \right) \psi (k) 
\nonumber \\
&&\quad+\sum_{j=1}^2
 \fourint{P} \kint \bar\psi (k + \half P) 
\left[ \sigma_j (P) + i\gamma_5 \lambda^a \pi_j^a (P)  \right]f_j(k_\perp ) 
 \psi (k - \half P) 
\nonumber \\
&&\quad- \frac{1}{2G}\sum_{j = 1}^2 \fourint{P} 
\left( \sigma_j (-P) \sigma_j (P) + \pi_j^a (-P) \pi_j^a (P) 
\right) .
\label{S_momentum}
\ea
Here it is understood that a cutoff in the 3--dimensional transverse
momentum is applied to the $k$--integral, as defined by the step
function of Eq.(\ref{F_2}).

\subsection{ Effective Lagrangian for $\pi$ and $\pi'$ mesons}
We now want to derive the effective Lagrangian describing physical
$\pi$ and $\pi'$ mesons. Integrating over the fermion fields in
Eq.(\ref{S_momentum}), one obtains the effective action of the
$\sigma_1 , \pi_1$-- and $\sigma_2 , \pi_2$--fields,
\ba
W[\sigma_1 , \pi_1 , \sigma_2 , \pi_2] &=& 
-\frac{1}{2G} \fourint{x} \ (\sigma_1^2 + \pi_1^{a\, 2} +
\sigma_2^2 + \pi_2^{a\, 2} ) \nonumber \\
&-& i N_c \; {\rm Tr}\, \log \left[ 
i \partialslash - m^0 + \sum_{j=1}^2 
(\sigma_j + i \gamma_5 \lambda^a \pi_j^a) f_j \right].
\label{W}
\ea
This expression is understood as a shorthand notation for expanding
in the meson fields.  In particular, we want to derive the free part
of the effective action for the $\pi_1$-- and $\pi_2$--fields,
\ba
W &=& W^{(0)} + W^{(2)}, \\
W^{(2)} &=& \frac12 \int\frac{d^4 P}{(2\pi )^4} 
\sum_{i, j = 1}^{2} \pi_i^a (P) K_{ij}^{ab} (P) \pi_j^b (P) ,
\label{W_2}
\ea
where  we restrict ourselves to timelike
momenta, $P^2 > 0$.  Before expanding in the $\pi_1$-- and
$\pi_2$--fields, we must determine the vacuum, {\em i.e.}, the mean
scalar field that arises in the dynamic breaking of chiral
symmetry.  The mean--field approximation corresponds to the leading
order of the $1/N_c$--expansion. The mean field is determined by the
set of equations
\ba
\frac{\delta W}{\delta\sigma_1} &=& - i N_c \; {\rm tr} \kint
\frac{1}{\rlap/k - m^0 + \sigma_1 + \sigma_2 f(k_\perp )}
- \frac{\sigma_1}{G} \; = \; 0 , \label{gap_1A}\\
\frac{\delta W}{\delta\sigma_2} &=& - i N_c \; {\rm tr} \kint
\frac{f(k_\perp )}{\rlap/k - m^0 + \sigma_1 + \sigma_2 f(k_\perp )}
- \frac{\sigma_2}{G} \; = \; 0 . \label{gap_2A}
\ea
Due to the transverse definition of the interaction,
Eq.(\ref{markov_yukawa}), the mean field inside a meson depends in a
trivial way on the direction of the meson 4--momentum, $P$. In the
following we consider these equations in the rest frame where
${\bf P} = 0, k_\perp = (0, \bfk )$ and $\Lambda_3$ is the usual
3--momentum cutoff.
\par
In general, the solution of Eqs.(\ref{gap_1A}) and (\ref{gap_2A}) would have
$\sigma_2\neq 0$, in which case the dynamically generated quark mass,
$-\sigma_1 - \sigma_2 f(\bfk ) + m^0$, becomes momentum--dependent.
However, if we choose the form factor, $f(\bfk )$, such that
\ba
-4 m I_1^f &\equiv&
 - i N_c \; {\rm tr} \kint
\frac{f(\bfk )}{\rlap/k - m} \;\; =\;\;
i 4 N_{\rm c} m \kint \frac{f(\bfk )}{m^2 - k^2} \;\; = \;\; 0 , 
\label{cond} \\
m &\equiv& - \sigma_1 + m^0 ,
\nonumber
\ea
then Eqs.(\ref{gap_1A}) and (\ref{gap_2A}) admit a solution with $\sigma_2 = 0$
and thus with a constant quark mass, $m = -\sigma_1 + m^0$. In this
case, Eq.(\ref{gap_1A}) reduces to the usual gap equation of the NJL
model,
\be
- 8 m I_1
\equiv - m i N_{\rm c} \kint \frac{1}{k^2 - m^2} 
\; = \; \frac{m^0 - m}{G}.
\label{gap_njl}
\ee
Obviously, the condition, Eq.(\ref{cond}), can be fulfilled by choosing an
appropriate value of the parameter $d$ defining the ``excited state''
form factor, Eq.(\ref{f}), for given values of $\Lambda_3$ and
$m$. Equation(\ref{cond}) expresses the invariance of the usual NJL
vacuum, $\sigma_1 = {\rm const.}$, with respect to variations in the
direction of $\sigma_2$. In the following, we shall consider the
vacuum as defined by Eqs.(\ref{cond}) and (\ref{gap_njl}), {\em i.e.}, we
work with the usual NJL vacuum. We emphasize that this choice is a
matter of convenience, not of principle. The qualitative results
below could equivalently be obtained with a different choice of  form
factors; however, in this case one should re-derive all vacuum
and ground--state meson properties with the momentum--dependent quark
mass. Preserving the NJL vacuum makes formulas below much more
transparent and allows us take the parameters fixed in the old
NJL model.

With the mean field determined by Eqs.(\ref{cond}) and (\ref{gap_njl}), we
now expand the action to quadratic order in the fields $\pi_1$ and
$\pi_2$.  The quadratic form $K_{ij}^{ab} (P)$, Eq.(\ref{W_2}), is
obtained as
\ba
K_{ij}^{ab} (P) &\equiv& \delta^{ab} K_{ij} (P) , \nonumber \\
K_{ij} (P) &=& -i N_{\rm c} \; {\rm tr}\, \kint \left[
\frac{1}{\kslash + \half\Pslash - m}
i\gamma_5 f_i
\frac{1}{\kslash - \half\Pslash - m} i \gamma_5 f_j
\right]  - \delta_{ij} \frac{1}{G} , 
\label{K_fullA}
\ea
A graphical representation of the loop integrals in Eq.(\ref{K_fullA})
is given in Fig.~\ref{loop1}.  The integral is evaluated by expanding in the
meson field momentum, $P$. To order $P^2$, one obtains
\ba
K_{11}(P) &=& Z_1 (P^2 - M_1^2 ), 
\hspace{2em} K_{22}(P) \;\; = \;\; Z_2 (P^2 - M_2^2 ) \nonumber \\
K_{12}(P) &=& K_{21}(P) \;\; = \;\;
\sqrt{Z_1 Z_2} \, \Gamma P^2 
\label{K_matrix}
\ea
where
\ba
Z_1 &=& 4 I_2, \hspace{2em} Z_2 \; = \; 4 I_2^{ff}, 
\label{I_12} \\
M_1^2 &=& Z_1^{-1}(-8I_1 + G^{-1}) \; = \; \frac{m^0}{Z_1 G m} , 
\label{m_1} \\
M_2^2 &=& Z_2^{-1}(-8I_1^{ff} + G^{-1}) , 
\label{m_2} \\
\Gamma &=& \frac{4}{\sqrt{Z_1 Z_2}} I_2^f .
\label{gamma}
\ea
Here, $I_n, I_n^f$, and $I_n^{ff}$ denote the usual loop integrals
arising in the momentum expansion of the NJL quark determinant, but
now with zero, one or two factors $f(k_\perp )$, Eq.(\ref{f}), in the
numerator. We may evaluate them in the rest frame, 
$k_\perp = (0, \bfk )$,
\be
I_n^{f..f} \equiv -i N_{\rm c} 
\kint \frac{f(\bfk )..f(\bfk )}{(m^2 - k^2)^n}.
\label{I_n}
\ee
The evaluation of these integrals with a 3--momentum cutoff is
described, {\em e.g.\/}, in ref.\cite{Ebert_93}. The integral over $k_0$
is taken by contour integration, and the remaining 3--dimensional
integral is regularized by the cutoff. Only the divergent parts are
kept; all finite parts are dropped. We point out that the momentum
expansion of the quark loop integrals, Eq.(\ref{K_fullA}), is an
essential part of this approach.  The NJL--model is understood here
as a model only for the lowest coefficients of the momentum expansion
of the quark loop, but not its full momentum dependence (singularities
{\em etc.}).
\par
Note that a mixing between the $\pi_1$ and $\pi_2$ fields occurs
only in the kinetic (${\cal O}(P^2 )$) terms of
Eq.(\ref{K_matrix}), but not in the mass terms. This is a direct
consequence of the definition of vacuum by Eqs.(\ref{cond}) and
(\ref{gap_njl}), which ensures that the quark loop with one form
factor has no $P^2$--independent part. The ``softness'' of the
$\pi_1$--$\pi_2$ mixing causes the $\pi_1$--field to decouple as
$P^2\rightarrow 0$. This property is crucial for the appearance of a
Goldstone boson in the chiral limit.
\par
To determine the physical $\pi$-- and $\pi'$--meson states, we have to
diagonalize the quadratic part of the action, Eq.(\ref{W_2}). If one
knew the full momentum dependence of the quadratic form,
Eq.(\ref{K_matrix}), the masses of  physical states would be given
as  zeros of the determinant of the quadratic form,
\be
\det K_{ij} (P^2 ) = 0, \hspace{2cm} P^2 \;\; = \;\; M_\pi^2 , \;
M_{\pi'}^2  .
\label{determinant}
\ee
This would be equivalent to the usual Bethe--Salpeter (on--shell)
description of bound states: the matrix $K_{ij}(P^2)$ is diagonalized
independently of the respective mass shells, 
$P^2 = M_\pi^2 , M_{\pi'}^2$ \cite{leyaouanc_84,weiss_93,gross}.  In
our approach, however, we know the quadratic form,
Eq.(\ref{K_matrix}), only as an expansion in $P^2$ at $P^2 = 0$. It
is clear that the determination of the masses according to
Eq.(\ref{determinant}) would be incompatible with the momentum
expansion, as the determinant involves ${\cal O}(P^4 )$--terms
neglected in Eq.(\ref{K_matrix}).  To be consistent with the
$P^2$--expansion, we must diagonalize the kinetic term and the mass
term in Eq.(\ref{W_2}) simultaneously, with a $P^2$--independent
transformation of the fields. Let us write Eq.(\ref{K_matrix}) in
the matrix form
\ba
K_{ij}(P^2 ) &=& \left(\begin{array}{cc}
Z_1 & \sqrt{Z_1 Z_2} \, \Gamma \\
\sqrt{Z_1 Z_2} \, \Gamma & Z_2 \end{array}\right) P^2 \;\; - \;\;
\left(\begin{array}{cc}
Z_1 M_1^2 & 0  \\ 0 & Z_2 M_2^2 \end{array}\right) .
\ea
The transformation that diagonalizes both the matrices here separately
is given by
\ba
\begin{array}{lcrcr}
\sqrt{Z_1} \pi_1^a &=& 
{\displaystyle \frac{\cos\phi}{\sqrt{{Z_\pi}}} \pi^a }
&+& {\displaystyle \frac{M_2}{M_1} \frac{\sin\phi}{\sqrt{Z_{\pi'}}} 
\pi^{\prime\, a}} , 
\\[.1cm] 
\sqrt{Z_2} \pi_2^a &=& {\displaystyle
\frac{M_1}{M_2}  \frac{\sin\phi}{\sqrt{Z_\pi}} \pi^a }
&-& {\displaystyle \frac{\cos\phi}{\sqrt{Z_{\pi'}}} 
\pi^{\prime\, a} } ,
\end{array}
\label{transform}
\ea
where
\ba
\tan 2\phi &=& 2 \Gamma \frac{M_1}{M_2} 
\left( 1 - \frac{M_1^2}{M_2^2} \right)^{-1} , 
\label{alpha} \\
Z_\pi &=& \cos^2\phi + \frac{M_1^2}{M_2^2} \sin^2\phi
+ 2 \Gamma \frac{M_1}{M_2} \cos\phi\sin\phi , \\
Z_{\pi'} &=& \cos^2\phi + \frac{M_2^2}{M_1^2} \sin^2\phi
- 2 \Gamma \frac{M_2}{M_1} \cos\phi\sin\phi .
\ea
In terms of the new fields, $\pi , \pi'$, the quadratic part of the
action, Eq.(\ref{W_2}), reads
\ba 
W^{(2)} &=& \half \fourint{P} \left[ 
\pi^a (-P) (P^2 - M_\pi^2) \pi^a (P)
+ \pi^{\prime\, a} (-P) (P^2 - M_{\pi'}^2) \pi^{\prime\, a} (P) 
\right] .
\label{S_2_phys}
\ea
Here,
\be
M_\pi^2 = \frac{M_1^2}{Z_\pi}, \hspace{2em}
M_{\pi'}^2  \;\; = \;\; \frac{M_2^2}{Z_{\pi'}} .
\label{mpi_phys}
\ee
The fields $\pi$ and $\pi'$ can thus be associated with physical
particles.
\par
Let us now consider the chiral limit, {\em i.e.}, a vanishing current
quark mass, $m^0 \rightarrow 0$. From Eqs.(\ref{I_12})--(\ref{gamma})
we see that this is equivalent to letting $M_1^2 \rightarrow 0$.
(Here and in the following, when discussing the dependence of
quantities on the current quark mass, $m^0$, we keep the constituent
quark mass fixed and assume the coupling constant, $G$, to be changed
in accordance with $m^0$, such that the gap equation,
Eq.(\ref{gap_njl}) remains fulfilled exactly. In this way, the loop
integrals and Eq.(\ref{cond}) remain unaffected by changes of the
current quark mass.)  Expanding Eqs.(\ref{mpi_phys}) in 
$M_1^2 \propto m^0$, one finds
\ba
M_\pi^2 &=& M_1^2 \; + \; {\cal O}(m_1^4 ), 
\label{mpi_chiral}
\\
M_{\pi'}^2 &=& \frac{M_2^2}{1 - \Gamma^2} 
\left[ 1 \; + \; \Gamma^2 \frac{M_1^2}{M_2^2}
 \; + \; {\cal O}(M_1^4 ) \right] .
\label{mpip_chiral}
\ea
Thus, in the chiral limit the effective Lagrangian,
Eq.(\ref{S_2_phys}), indeed describes a massless Goldstone pion,
$\pi$, and a massive particle, $\pi'$.  Furthermore, in the chiral
limit the transformation of the fields, Eq.(\ref{transform}), becomes
\ba
\sqrt{Z_1} \pi_1^a &=& 
{\displaystyle \left( 1 - \Gamma^2 \frac{M_1^2}{M_2^2} \right) 
\pi^a}
+ {\displaystyle \frac{\Gamma}{\sqrt{1 - \Gamma^2}} \left(
1 + (1 - \Gamma^2 ) \frac{M_1^2}{M_2^2} \right) 
\pi^{\prime\, a}} , \nonumber\\
\sqrt{Z_2} \pi_2^a &=& 
{\displaystyle \Gamma \frac{M_1^2}{M_2^2} \pi^a }
- {\displaystyle \frac{1}{\sqrt{1 - \Gamma^2}} 
\pi^{\prime\, a} } . 
\label{transform_chiral}
\ea
At $M_1^2 = 0$ one observes that $\pi$ has only a component along
$\pi_1$. This is a consequence of the fact that the $\pi_1$--$\pi_2$
coupling in the original Lagrangian, Eq.(\ref{K_matrix}), is of order
$P^2$.  We remark that, although we have chosen to work with the
particular choice of excited--state form factor, Eq.(\ref{cond}), the
occurrence of a Goldstone boson in the chiral limit in Eq.(\ref{W}) is
general and does not depend on this choice.  This may easily be
established by using the general gap equations, Eqs.(\ref{gap_1A}) and
(\ref{gap_2A}), together with Eq.(\ref{K_fullA}).

\subsection{The axial current}
To describe the leptonic decays of the $\pi$ and $\pi'$
mesons, we need the axial current operator.  Since our effective
action contains, besides the pion, a field describing an ``excited
state'' with the same quantum numbers, it is clear that the axial
current of our model is, in general, not carried exclusively by the
$\pi$ field, and is thus not given by the standard PCAC formula.
Thus, we must determine the conserved axial current of our model,
including the contribution of $\pi'$, from first principles.

In general, the construction of the conserved current in a theory
with nonlocal (momentum--dependent) interactions is a difficult
task. This problem has been studied extensively in the framework of
the Bethe--Salpeter equation \cite{riska_87} and various
3--dimensional reductions of it such as the quasipotential and the
on--shell reduction \cite{gross_current}. In these approaches, the
derivation of the current is achieved by ``gauging'' all possible
momentum dependences of the interaction through minimal substitution,
a rather cumbersome procedure in practice. In contrast, in a
Lagrangian field theory, a simple method exists to derive conserved
currents, the so--called Gell--Mann and Levy method \cite{gell-mann},
based on the Noether theorem.  In this approach, the current is
obtained as the variation of the lagrgangian with respect to the
derivative of a space--time dependent symmetry transformation of the
fields. We now show that a suitable generalization of this technique
can be employed to derive the conserved axial current of our model
with quark--meson form factors depending on the transverse momentum.

To derive the axial current, we start at the quark level.  The isovector
axial current is the Noether current corresponding to infinitesimal
chiral rotations of the quark fields,
\be
\psi (x) \rightarrow \left( 1 - i \varepsilon^a \half \lambda^a 
\gamma_5 \right) \psi (x) .
\label{chiral_psi}
\ee
Following the usual procedure, we consider the parameter of this
transformation to be space--time dependent, 
$\varepsilon^a \equiv \varepsilon^a (x)$. However, this dependence
should not be completely arbitrary. To describe the decays of $\pi$ and
$\pi'$ mesons, it is sufficient to know the component of the axial
current parallel to the meson 4--momentum, $P$. It is easy to see
that this component is obtained from chiral rotations whose parameter
depends only on the longitudinal part of the coordinate
\be
\varepsilon^a (x) \rightarrow \varepsilon^a (x_{\vert\vert}),
\hspace{2cm}
x_{\vert\vert} \;\; \equiv \;\; \frac{x \cdot P}{\sqrt{P^2}}, 
\label{eps_x}
\ee
since $\partial_\mu \varepsilon^a (x_{\vert\vert}) \propto P_\mu$.
In other words, transformations of the form Eq.(\ref{eps_x}) describe
a transfer of the longitudinal momentum to the meson, but not of the
transverse momentum. This has an important consequence that the
chiral transformation does not change the direction of transversality
of the meson--quark interaction, {\em cf.}\ Eq.(\ref{markov_yukawa}).
When passing to the bosonized representation, Eq.(\ref{S_sep}), the
transformation of the $\pi_1 , \sigma_1$-- and 
$\pi_2 , \sigma_2$--fields induced by Eqs.(\ref{chiral_psi}) and
(\ref{eps_x}) is therefore of the form
\ba
\begin{array}{lclcl}
\pi^a_i (x) &\rightarrow& \pi^a_i (x) &+& 
\varepsilon^a (x_{\vert\vert}) \, \sigma_i (x) , \\
\sigma_i (x) &\rightarrow& \sigma_i (x) &-& 
\varepsilon^a (x_{\vert\vert}) \, \pi^a_i (x) .
\end{array}
\hspace{1.5cm} (i = 1, 2)
\label{chiral_pi}
\ea
This follows from the fact that, for a fixed direction of $P$, the
vertex, Eq.(\ref{markov_yukawa}), describes an instantaneous
interaction in $x_{\vert\vert}$. Thus, the special chiral rotation,
Eq.(\ref{eps_x}), does not mix the components of  meson fields
coupled to quarks with different form factors.
\par
With the transformation of the chiral fields given by
Eqs.(\ref{chiral_pi}), the construction of the axial current proceeds
exactly as in the usual linear sigma model. We write the variation of
the effective action, Eq.(\ref{W}), in the momentum representation,
\be
\delta W = \fourint{Q} \varepsilon^a (Q) D^a (Q) ,
\ee
where $\varepsilon^a (Q) = \tilde\varepsilon^a (Q_{\vert\vert})
\delta^{(3)} (Q_\perp )$ is the Fourier transform of the
transformation, Eq.(\ref{eps_x}), and $D^a (Q)$ is a function of the
fields $\sigma_i , \pi_i , i = 1, \ldots 2$, given in the form of a
quark loop integral,
\ba
D^a (Q)
&=& -i N_c \; {\rm tr} \fourint{k} 
\left[ \frac{1}{\kslash - m}\delta^{ab}
+ \frac{1}{\kslash - \frac12 \Qslash - m} i\gamma_5 \lambda^a
\frac{1}{\kslash + \frac12 \Qslash - m} i\gamma_5 \lambda^b 
\sigma_1 \right] 
\nonumber \\
&& \times (\pi_1^b (Q) + f(k_\perp ) \pi_2^b (Q) ) .
\ea
Here we have used that $\sigma_2 = 0$ in the vacuum,
Eq.(\ref{cond}). Expanding now in the momentum $Q$, making use of
Eq.(\ref{cond}) and the gap equation, Eq.(\ref{gap_njl}), and setting
$\sigma_1 = -m$ (it is sufficient to consider the symmetric limit,
$m^0 = 0$), we get
\ba
D^a (Q)
&=& -Q^2 m \left[ 4 I_2 \pi_1^a (Q) + 4 I_2^f \pi_2^a (Q) \right] 
\nonumber \\
&=& -Q^2 m \left[ Z_1 \pi_1^a (Q) + \sqrt{Z_1 Z_2} \Gamma
\pi_2^a (Q) \right] . 
\label{D_result}
\ea
The fact that $D^a (Q^2 )$ is proportional to $Q^2$ is a consequence
of the chiral symmetry of the effective action, Eq.(\ref{W}). Due to
this property, $D^a (Q^2 )$ can be regarded as the divergence of a
conserved current,
\be
A_\mu^a (Q) = Q_\mu m \left[ 
Z_1 \pi_1^a (Q) + \sqrt{Z_1 Z_2} \Gamma \pi_2^a (Q) \right] .
\label{axial_current}
\ee
Equation (\ref{axial_current}) is the conserved axial current of our model.
It is of the usual ``PCAC'' form, but contains also a contribution of
the $\pi_2$ field. The above derivation was rather formal. However,
the result can be understood in simple terms, as is shown in Fig.~\ref{pi-a}:
Both the $\pi_1$ and $\pi_2$--fields couple to the local axial current
of the quark field through quark loops; the $\pi_2$--field enters the
loop with a form factor, $f(k_\perp )$. The necessity to pull out a
factor of the meson field momentum (derivative) means that only the
${\cal O}(P^2)$--parts of the loop integrals, $I_2$ and $I_2^f$,
survive, {\em cf.}\ Eq.(\ref{I_n}). Chiral symmetry ensures that the
corresponding diagrams for the divergence of the current have no
$P^2$--independent part.

The results of this Subsection are an example for the technical
simplifications of working with separable quark interactions. The
fact that they can be bosonized by local meson fields makes it
possible to apply methods of local field theory, such as the Noether
theorem, to the meson effective  action. Furthermore, we note that the
covariant (transverse) definition of the 3--dimensional quark
interaction, Eq.(\ref{markov_yukawa}), is crucial for obtaining a
consistent axial current. In particular, with this formulation there
is no ambiguity with different definitions of the pion decay constant
like with non--covariant 3-dimensional interactions
\cite{leyaouanc_84}.
\subsection{ The weak decay constants of  $\pi$ and $\pi'$ mesons}
We now use the axial current derived in the previous Subsection to
evaluate the weak decay constants of physical $\pi$ and $\pi'$
mesons. They are defined by the matrix element of the divergence of
the axial current between meson states and  vacuum,
\ba
\langle 0 | \partial^\mu A_\mu^a | \pi^b \rangle &=&
M_\pi^2 F_\pi \delta^{ab} ,
\label{decay_pi} \\
\langle 0 | \partial^\mu A_\mu^a | \pi^{\prime\, b} \rangle &=&
M_{\pi'}^2 F_{\pi'} \delta^{ab}
\label{decay_pipr} .
\ea
In terms of the physical fields, $\pi$ and $\pi'$, the axial current
takes the form
\be
A_\mu^a = P_\mu m \sqrt{Z_1} \left( \pi^a 
\; + \; \Gamma \sqrt{1 - \Gamma^2} \, 
\frac{M_1^2}{M_2^2} \, \pi^{\prime\, a} \right) 
\; + \; {\cal O}(M_1^4 ).
\ee 
Here, we substituted the transformation of the fields,
Eq.(\ref{transform_chiral}), into Eq.(\ref{axial_current}).  The decay
constants of the physical $\pi$ and $\pi'$ states are thus given by
\ba
F_\pi &=& \sqrt{Z_1} m \; + \; {\cal O}(M_1^4 ), 
\label{fpi} \\
F_{\pi'} &=& \sqrt{Z_1} m \, \Gamma \sqrt{1 - \Gamma^2} \, 
\frac{M_1^2}{M_2^2} \; + \; {\cal O}(M_1^4 ) .
\label{fpip}
\ea
The corrections to $F_\pi$ for excited states are
of order $M_\pi^4$. Thus, within our accuracy, $F_\pi$ is identical with
the value obtained by the usual NJL model, $\sqrt{Z_1} m$, which 
follows from the Goldberger--Treiman relation at the quark 
level \cite{volkov_83}.  On the other hand, the $\pi'$ decay constant
vanishes in the chiral limit $m^0 \sim M_1^2 \rightarrow 0$, as
expected. We stress that for this property to hold, it is essential to
consider the full axial current, Eq.(\ref{axial_current}), including
the contribution of the $\pi_2$--component. As can be seen from
Eqs.(\ref{transform_chiral}) and (\ref{axial_current}), the
standard PCAC formula $A_\mu^a \propto \partial_\mu \pi_1^a$ would
lead to a non-vanishing result for $F_{\pi'}$ in the chiral limit.

The ratio of the $\pi'$ to $\pi$ decay constants can directly be
expressed in terms of the physical $\pi$ and $\pi'$ masses.  From
Eqs.(\ref{fpi}) and (\ref{fpip}) one obtains, using Eqs.~(\ref{mpi_chiral}) and
(\ref{mpip_chiral}),
\be
\frac{F_{\pi'}}{F_\pi} = \Gamma \sqrt{1 - \Gamma^2} \, 
\frac{M_1^2}{M_2^2}
\;\; = \;\; \frac{\Gamma}{\sqrt{1 - \Gamma^2}} \,
\frac{M_\pi^2}{M_{\pi'}^2}.
\label{ratio}
\ee
This is precisely the dependence derived from current
algebra considerations in the general ``extended PCAC'' framework
\cite{dominguez}.  We note that the same behavior of $F_{\pi'}$ in
the chiral limit is found in models describing chiral symmetry
breaking by nonlocal interactions \cite{leyaouanc_84,weiss_93}.
\par
The effective Lagrangian  in a compact way illustrates  different
consequences of axial current conservation for the pion and its
excited state. Both matrix elements of $\partial_\mu A^\mu$,
Eq.(\ref{decay_pi}) and Eq.(\ref{decay_pipr}), must vanish for 
$m^0 \rightarrow 0$.  The pion matrix element, Eq.(\ref{decay_pi}),
does so by $M_\pi^2 \rightarrow 0$, with $F_\pi$ remaining finite,
while for the excited pion matrix element the opposite takes place,
$F_{\pi'} \rightarrow 0$ with $M_{\pi'}$ remaining finite.

\subsection{ Numerical estimates and conclusions }
We can now  numerically estimate the excited pion decay constant,
$F_{\pi'}$, in this model. We take the value of the constituent quark
mass  $m = 300\,{\rm MeV}$ and fix the 3--momentum cutoff at
$\Lambda_3 = 671 \,{\rm MeV}$ by fitting the normal pion decay
constant $F_\pi = 93\, {\rm MeV}$ in the chiral limit, as in the
usual NJL model without excited states, {\em cf.} \cite{Ebert_93}.
With these parameters one obtains the standard value of the quark
condensate, $\langle \bar q q \rangle = - (253 \, {\rm MeV})^3$, and
$G = 0.82\, m^{-2} = 9.1\, {\rm GeV}^{-2}, \; m^0 = 5.1\, {\rm MeV}$.
With the constituent quark mass and cutoff fixed, we can determine
the parameter $d$ of the ``excited--state'' form factor,
Eq.(\ref{f}), from the condition Eq.(\ref{cond}).  We find%
\footnote{
All parameters will be different when in Section 3 we consider 
a realistic version of this model. However, the ratio $d/\Lambda_3$ will
be near 2 (its limit as $\Lambda\to\infty$) and change slightly.
} 
$d = -1.83 \, \Lambda_3^{-2} = -4.06\, {\rm GeV}^{-2}$, corresponding
to a form factor $f(k_\perp )$ with a radial node in the range 
$0 \leq | k_\perp | \leq \Lambda_3$.  With this value we determine
the $\pi_1$--$\pi_2$ mixing coefficient, $\Gamma$, Eq.(\ref{gamma}),
as
\be
\Gamma = 0.41.
\ee
Note that $\Gamma$ is independent of the normalization of the form
factor $f(k_\perp )$, Eq.(\ref{f}). In fact, the parameter $c$ enters
only into the mass of the $\pi'$ meson, {\em cf.}\ Eqs.(\ref{m_2}) and
(\ref{mpip_chiral}); we should not  determine its value since the
result can directly be expressed in terms of $M_{\pi'}$. Thus,
Eq.(\ref{ratio}) gives
\be
\frac{F_{\pi'}}{F_\pi} = 0.45 \frac{M_\pi^2}{M_{\pi'}^2}.
\ee
For the standard value of the $\pi'$ mass, 
$M_{\pi'} = 1300\mev$, this comes to 
$F_{\pi'} = 0.48\mev$.
The excited pion leptonic decay
constant is thus very small, which is a consequence of chiral
symmetry. Note that, as opposed to the qualitative results discussed
above, the numerical values here depend on the choice of  form factor,
(see Eq.(\ref{cond})), and should thus be regarded as a rough estimate.

We remark that the numerical values of the ratio $F_{\pi'}/F_\pi$
obtained here are comparable to those found in chirally symmetric
potential models \cite{weiss_93}. However, models describing chiral
symmetry breaking by a vector--type confining potential (linear or
oscillator) usually underestimate the normal pion decay constant by an
order of magnitude \cite{leyaouanc_84}. Such models should include a
short--range interaction (NJL--type) which is mostly responsible for
chiral symmetry breaking.

The small value of $F_{\pi'}$ does not imply a small width of the
$\pi'$ resonance, since it can decay hadronically, {\em e.g.\/}, into
$3\pi$ or $\rho\pi$. Such hadronic decays will  be investigated in
Section 4.

In conclusion, we  outlined a simple framework for including
radial excitations in an effective Lagrangian description of
mesons. The Lagrangian obtained by bosonization of an NJL--model with
separable interactions exhibits all qualitative properties expected
on general grounds: a Goldstone pion with a finite decay constant,
and a massive ``excited state'' with a vanishing decay constant in the
chiral limit. Our model shows in a simple way how chiral symmetry
protects the pion from modifications by excited states, which in turn
influences the excited states' contribution to the axial
current. These features are general and do not depend on a particular
choice of the quark--meson form factor.  Furthermore, they are preserved
if the derivative expansion of the quark loop is carried to higher
orders.

In the investigations described here we strictly kept to
an effective Lagrangian approach, where the coupling constants and
field transformations are defined at zero momentum. We have no way to
check the quantitative reliability of this approximation for radially
excited states in the region of $\sim 1\,{\rm GeV}$, {\em i.e.}, to
estimate the momentum dependence of the coupling constants, within
the present model. (For a general discussion of the range of
applicability of effective Lagrangians, see \cite{jaffe_92}.)  This
question can be addressed to generalizations of the NJL model with
quark confinement, which in principle allow both a zero--momentum and
an on--shell description of bound states. Recently, first
steps were undertaken to investigate the full momentum dependence of
correlation functions in  an approach of that kind \cite{celenza}.

\section{$U(3)\times U(3)$ model.}

\subsection{$U(3)\times U(3)$ chiral Lagrangian with excited meson states
and 't Hooft interaction}

This Section is devoted to a realistic $U(3)\times U(3)$ version
of the NJL model with nonlocal four-quark interaction (see 
refs.~\cite{volk_97,volk_99A,volk_99B}).

We use a  nonlocal separable four-quark interaction
of the current-current form  
which admits nonlocal vertices (form
factors) in the quark currents and a pure local six-quark 't Hooft
interaction 
\cite{klev,volk98}:
\ba
     {\cal L}(\bar q, q) &=&
     \int\! d^4x\; \bar q(x)
     (i \partialslash -m^0) q(x)+
     {\cal L}^{(4)}_{\rm int}+
     {\cal L}^{(6)}_{\rm int},  \label{lag}\\
     {\cal L}^{(4)}_{\rm int} &=&
     \frac{G}{2}\int\! d^4x\sum^{9}_{a=1}\sum^{N}_{i=1}
     [j_{S,i}^a(x) j_{S,i}^a(x)+
     j_{P,i}^a(x) j_{P,i}^a(x)]\nonumber\\
     &&-\frac{G_V}{2}\int\! d^4x\sum^{9}_{a=1}\sum^{N}_{i=1}
     [j_{V,i}^{a,\;\mu}(x) j_{V,i,\; \mu}^a(x)+
     j_{A,i}^{a,\;\mu}(x) j_{A,i,\;\mu}^a(x)],\\
     {\cal L}^{(6)}_{\rm int}&=&-K \left[\det
     \left[\bar q (1+\gamma_5)q\right]+
     \det\left[\bar q (1-\gamma_5)q\right]
     \right]
\ea
where ${\cal L}^{(4)}_{\rm int}$ is the $U(3)\times U(3)$ chirally
symmetric four-quark interaction Lagrangian and 
${\cal L}^{(6)}_{\rm int}$ contains the symmetry breaking 't Hooft terms.
Here, $m^0$ is the current quark mass matrix 
$m^0=\diag (m_u^0,m_d^0,m_s^0)$
($m_u^0\approx m_d^0$) and
$j^a_{U,i}$ with $U=(S,P,V,A)$ denotes the scalar, pseudoscalar, vector, and
axial-vector  quark currents
\be
     j^a_{S(P),i}(x)=
     \int\! d^4x_1 d^4x_2\; \bar q(x_1)
     F^a_{S(P),i }(x;x_1,x_2) q(x_2),
\ee
\be
     j^{a,\;\mu}_{V(A),i}(x)=
     \int\! d^4x_1 d^4x_2\; \bar q(x_1)
     F^{a,\;\mu}_{V(A),i }(x;x_1,x_2) q(x_2)
\ee
where $ F^a_{S(P),i}(x;x_1,x_2)$ are the scalar (pseudoscalar) and
$ F^{a,\;\mu}_{V(A),i}(x;x_1,x_2)$ the vector and axial-vector
nonlocal quark vertices. The index $a=1,\ldots,9$ denotes  the 
basis elements $\tau^a$ of $U(3)$ flavor group. 
Our choice is slightly different from the Gell-Mann $\lambda$ matrices
\ba
     &&{\tau}_i={\lambda}_i ~~~ (i=1,...,7),~~~
     \tau_8 = (
     \lambda_0 + \lambda_8)/{\sqrt 3}=
	\left(\begin{array}{ccc}
	    1 & 0 & 0\\
	    0 & 1 & 0\\
	    0 & 0 & 0\\
	\end{array}\right)
	,\nonumber\\
&&   \tau_9  = (-\lambda_0 +
     {\sqrt 2}\lambda_8)/{\sqrt 3}=
	\left(\begin{array}{ccc}
	    0 & 0 & 0\\
	    0 & 0 & 0\\
	    0 & 0 & -\sqrt{2}\\
	\end{array}\right)
, \label{DefTau} 
\ea 
but this choice is more convenient when a singlet-octet mixing appears due to 
the 't Hooft  terms. 

In the original formulation of the NJL model with 't Hooft interaction, 
the 't Hooft terms are represented by six-fermion vertices. In this form
the Lagrangian is not ready for the bosonization procedure,  we
should proceed to.  An appropriate way to circumvent this drawback is
to come to an equivalent form of the quark Lagrangian that contains
only four-quark vertices as it was done, {\it e.g.\/}, in refs. \cite{dmitr,klev}. 
Therein, the effective four-quark interaction is deduced by integrating
out a quark loop at each six-quark vertex. 
Thus, from ${\cal L}^{(6)}_{\rm int}$ the four-quark part 
${\cal L}^{(4)}_{\rm int}$ acquires  an additional contribution which in
the one-loop approximation looks as follows:
\ba
L^{(4)}_{\rm tH}&=&4K\int d^4x\left\{
	\sum_{a=1}^3m_s I_1(m_s)\left[(\bar q(x) i\gamma_5\tau^a q(x))^2 -
	(\bar q(x)\tau^a q(x))^2\right]\right.\nonumber\\
&& 
	+\sum_{a=4}^7m_u I_1(m_u)\left[(\bar q(x)i\gamma_5\tau^a q(x))^2 -
	(\bar q(x)\tau^a q(x))^2\right]\nonumber\\
&&	
	+m_s I_1(m_s)\left[(\bar q(x)\tau^8 q(x))^2 -
	(\bar q(x)i \gamma_5\tau^8 q(x))^2\right]\nonumber\\
&&		
	-2\sqrt{2} m_u I_1(m_u)\left[(\bar q(x)\tau^8 q(x))(\bar q(x)\tau^9 q(x)) 
\right.\nonumber\\
&&\left.
	-(\bar q(x)i \gamma_5\tau^8 q(x))(\bar q(x)i \gamma_5\tau^9 q(x))\right]
	\Biggr\}.
	\label{tHooft4q}
	\ea

In our model the 't Hooft interaction is local with respect to its instantaneous
origin.
Finally, we have%
\footnote{
It should be noted  that SBCS
is already taken into account in the effective four-fermion vertices. 
Therefore, the effective four fermion Lagrangian is no longer chirally invariant.
However, in its original form the chiral invariance is present if we exclude
't Hooft terms. This fact has some consequences which we use later,
for instance, we choose the same form factors both for  scalars and pseudoscalars.
}:
\ba
     {\cal L}(\bar q, q) &=&
     \int\! d^4x\; \bar q(x)
     (i \partialslash -m^0) q(x)\nonumber\\
     &&+\frac{1}{2}\int\! d^4x\sum^{9}_{a=1}\sum^{9}_{b=1}
     [G^{(-)}_{ab}j_{S,1}^a(x) j_{S,1}^b(x)+
     G^{(+)}_{ab}j_{P,1}^a(x) j_{P,1}^b(x)]\nonumber\\ 
     &&+\frac{G}{2}\int\! d^4x\sum^{9}_{a=1}\sum^{N}_{i=2}
     [j_{S,i}^a(x) j_{S,i}^a(x)+
     j_{P,i}^a(x) j_{P,i}^a(x)]\nonumber\\
     &&-\frac{G_V}{2}\int\! d^4x\sum^{9}_{a=1}\sum^{N}_{i=1}
     [j_{V,i}^{a,\;\mu}(x) j_{V,i,\; \mu}^a(x)+
     j_{A,i}^{a,\;\mu}(x) j_{A,i,\;\mu}^a(x)].
\ea
where
\ba
&&   G_{11}^{(\pm)}=G_{22}^{(\pm)}=G_{33}^{(\pm)}=
     G \pm 4Km_sI_1(m_s), \nonumber \\
&&   G_{44}^{(\pm)}=G_{55}^{(\pm)}=G_{66}^{(\pm)}=
     G_{77}^{(\pm)}= G \pm 4Km_uI_1(m_u),
     \nonumber \\
&&   G_{88}^{(\pm)}= G \mp 4Km_sI_1(m_s), ~~~
     G_{99}^{(\pm)}= G, ~~~
     G_{89}^{(\pm)}= G_{98}^{(\pm)}=\pm 4{\sqrt 2}Km_uI_1(m_u),\nonumber\\
&&   G_{ab}=0\quad (a\not=b,\; a,b=1,\ldots,7).
	\label{DefG}
\ea

 The model thus formulated can be bosonized in a standard way by
introducing auxiliary boson fields $\sigma_i^a(x),\varphi_i^a(x),V^\mu_i(x),A^\mu_i(x)$ 
with quantum numbers of
the quark currents $j^a_{S(P),i}(x)$ and $j^{a,\;\mu}_{V(A),i}$, and
then integrating over the quark degrees of freedom. The result is 
a meson effective  Lagrangian which, after all, is a functional of 
scalar, pseudoscalar, vector, and axial-vector meson
fields. 
In the case of an ordinary (local) NJL model, this procedure would 
give us the well-known linear realization of the chiral Lagrangian.
When  original four-quark vertices of the separable type
contain form factors, the bosonization gives rise to a meson effective
 Lagrangian for the ground state and a number (in general infinite) of
radially excited meson fields. These fields have the same quantum 
numbers and therefore should be interpreted as ``radial'' excitations.

The effective four-quark representation of the Lagrangian with 
't Hooft interaction requires careful treatment. It is not equivalent
to the original form in all aspects. For example, the gap equations 
derived from the effective four-quark form of the Lagrangian do
not reproduce those obtained from the original form (with six-quark vertices).
A kind of double counting takes place here, which leads to wrong gap equations
(for a correct derivation of gap equations, see~\cite{klev}). 
But for the mass spectra and meson-meson coupling constants
in the one-loop approximation, everything works well.

In the one-loop approximation, the bosonized Lagrangian has the
following form:
\ba
&&L_{\rm bos}(\bar q, q; \sigma, \varphi, V, A) = \int d^4 x_1
\int d^4 x_2~ \bar q (x_1 ) \left[ \left( i \partialslash_{x_2}
- m^0 \right) \delta (x_1 - x_2 )\right.      \nonumber \\
&&+ \int d^4 x  \sum_{i = 1}^N\sum_{a=1}^9
\left( \sigma^a_i (x) F^a_{\sigma , i} (x; x_1, x_2 ) +
\varphi_i^a (x) F_{\varphi , i}^a (x; x_1, x_2)  \right. \nonumber \\
&&\left.\left.
+ V_{i,\mu}^{a} (x) F_{V , i}^{a,\mu} (x; x_1, x_2) +
A_{i,\mu}^{a} (x) F_{A , i}^{a,\mu} (x; x_1, x_2) \right) \right] q (x_2 )
\nonumber \\
&&- \sum_{a=1}^9\int d^4 x 
\left[ \frac{1}{2} \left(\left(G^{(-)}\right)^{-1}_{ab} \sigma_1^{a} (x)\sigma_1^{b} (x) +
\left(G^{(+)}\right)^{-1}_{ab}\varphi_1^{a} (x)\varphi_1^{b} (x) \right)\right.\nonumber\\ 
&&\left.-\frac{1}{2G_V} \left(\left( V_1^{a,\mu} (x)\right)^2 + 
\left(A_1^{a,\mu} (x)\right)^2
\right) \right]\nonumber\\
&&- \int d^4 x \sum_{i = 2}^N
\left[ \frac{1}{2G} \left(\left( \sigma_i^{a} (x)\right)^2 +
\left(\phi_i^{a} (x)\right)^2 \right)
- \frac{1}{2G_V} \left(\left( V_i^{a,\mu} (x)\right)^2 + \left(A_i^{a,\mu} (x)\right)^2
\right) \right].
\label{L_sep}
\ea

This Lagrangian describes a system of local meson fields,
$\sigma_i^a (x)$, $\phi_i^a (x)$, $V^{a,\mu}_i (x)$, $A^{a,\mu}_i (x)$,
$i = 1, \ldots N$, which interact with
 quarks through nonlocal vertices. These fields
are not yet to be associated with physical particles,
to be obtained after determining the vacuum and
diagonalizing the meson effective Lagrangian.

In general, the model admits as many excited states as one wishes.
But for a realistic description of very heavy mesons ($2 \gev$ and more) the model
seems not reliable because it is constructed for low energies. 
So we intended here to consider a minimal version of the model,
restricting ourselves to $N=2$, which is necessary for 
the description of  ground states and first radial excitations of mesons.

To describe the ground and  first radially excited states 
of mesons, we take the form factors in the momentum representation as follows:
\be
\begin{array}{ll}
     F_{S,j}^a({\bf k})=\tau^a
     f^a_{\sigma,j},\quad & F_{P,j}^a=
     i\gamma_5 \tau^a f^a_{\varphi,j},
\end{array}
\ee
\be
\begin{array}{ll}
     F_{V,j}^{a,\;\mu}({\bf k})=\gamma^\mu\tau^a
     f^a_{V,j},\quad & F_{A,j}^{a,\;\mu}=
     \gamma_5\gamma^\mu \tau^a f^a_{A,j},
\end{array}
\ee

\be
     f^a_{U,1}\equiv 1,\quad f^a_{U,2}\equiv f_a^{U}({\bf k})=c_a^{U}(1+d_a {\bf k}^2),\label{fDef}
\ee
where $U=(\sigma,\varphi,V,A)$.
Here, we consider the form factors in
the rest frame of mesons (see Section 2).
After bosonization in the one-loop approximation, we get 
\ba
&&L_{\rm bos}(\sigma, \varphi, V, A) = \nonumber\\
&&\quad- \sum_{a,b=1}^9\int d^4 x 
\left[ \frac{1}{2} \left(\left(G^{(-)}\right)^{-1}_{ab} 
\bar\sigma_1^{a} (x)\bar\sigma_1^{b} (x) +
\left(G^{(+)}\right)^{-1}_{ab}\varphi_1^{a} (x)\varphi_1^{b} (x) \right)\right.\nonumber\\ 
&&\quad\left.-\frac{1}{2G_V} \left( \left(V_1^{a,\mu} (x)\right)^2 + 
\left(A_1^{a,\mu} (x)\right)^2
\right) \right]\nonumber\\
&&\quad- \sum_{a=1}^9\int d^4 x 
\left[ \frac{1}{2G} \left( \left(\sigma_2^{a} (x)\right)^2 +
\left(\phi_2^{a} (x)\right)^2 \right)
- \frac{1}{2G_V} \left( \left(V_2^{a,\mu} (x)\right)^2 +\left( A_2^{a,\mu} (x)\right)^2
\right) \right]\nonumber\\
&&\quad-i {\rm Tr}\ln \biggl[1+\frac{1}{i\!\!\not\!\partial-m}\sum_{j=1}^2 \sum_{a=1}^9
( \sigma^a_j+ \varphi_j^a   +V_j^{a,\mu} \gamma_\mu +
A_j^{a,\mu} \gamma_5\gamma_\mu) f_{j}^{a}\tau_a
   \biggr] 
\label{L_bos}
\ea

At the beginning of this Section, we have already mentioned
that there is a danger of double counting when deriving
gap equations. The double counting surely takes place if
one tries to obtain the gap equations by  na\"\i vely varying
the Lagrangian (\ref{L_bos}) over $\sigma^a_1$. However,
correct equations for $\sigma^a_2$ can be obtained in
this way. It is due to the fact that the 't Hooft
interaction is local. 

The gap equations for 
$\sigma^a_1$ can be deduced from the Dyson-Schwinger equation.
We will not discuss the details of finding its solution 
but refer the reader, {\it e.~g.\/}, 
to paper \cite{klev}. Here we present just the result
that is a slight modification of the equations obtained
in ref.~\cite{klev}. 
\ba
m^0_u&=&m_u[1-8G_{88}^{(-)}(I_1(m_u)+I_1^{f_{uu}}(m_u)f^8_2)],\label{gap1}\\
m^0_s&=&m_s[1-8G_{99}^{(-)}(I_1(m_s)+I_1^{f_{ss}}(m_s)f^9_2)].
\label{gap2}
\ea 

There $m^0_a$ and $m_a$ ($a=u,d,s$) are the current and constituent
quark masses, respectively.
The difference between Eqs.(\ref{gap1}),(\ref{gap2}) and those given in ref.\cite{klev}
is the presence of $I_1^f(m_u)$, tadpoles with form factors 
absent in local NJL.

The constituent quark masses appear, as usual, due to non-zero vacuum expectations
of $\sigma^a$, according to the equations 
\be
\langle\bar\sigma^8\rangle_0=m^0_u-m_u,
\qquad \langle\bar\sigma^9\rangle_0=m^0_s-m_s.
\ee
We  use them in the gap equations for  excited meson states.
The fields $\bar\sigma^a$ require redefinition which consists in
subtracting their vacuum expectation values:
\be
\sigma^8=\bar\sigma^8-\langle\bar\sigma^8\rangle_0,\qquad 
\sigma^9=\bar\sigma^9-\langle\bar\sigma^9\rangle_0.
\ee

Now we stop discussing the gap equation for the ground fields and
turn our attention to those for radially excited meson states. 
As it was said above,
the correct gap equations for radially excited meson states
can be obtained by calculating
the first derivative of 
Lagrangian (\ref{L_bos}) with respect to $\sigma^a_2$, which  gives 
\be
\langle\frac{\delta L}{\delta\sigma^a_2}\rangle_0 = - i N_c \; {\rm tr} \kint
\frac{f^a({\bf k})}{( \rlap/k - m + \langle\sigma^a_2\rangle_0\tau_a f^a({\bf k}))}
- \frac{\langle\sigma^a_2\rangle_0}{G} \; = \; 0.
\label{gap_1}
\ee
This equation always admits the trivial solution $\langle\sigma^a_2\rangle=0$.
Despite  the fact that nontrivial solutions are possible,
 we assume  that the vacuum expectations
for  radially excited meson states are equal to zero and 
therefore do not change the quark condensate. 
Thus, we obtain the condition
\be
- i N_c \; {\rm tr} \kint
\frac{f^a({\bf k})}{( \rlap/k - m  )} = \; 0.
\label{gap_2}
\ee
Equation (\ref{gap_2}) is written in the matrix form. In the isotopic symmetry,
Eq.(\ref{gap_2}) gives two conditions on the form factors $f^a({\bf k})$
which can be written in our notation as follows:
\ba
I_1^{f_{uu}}(m_u)&=&0\label{condition1},\\
I_1^{f_{ss}}(m_s)&=&0. \label{condition2}
\ea
These conditions  essentially simplify the calculation of the meson
mass spectra. In particular, they provide a diagonal form
for the $(\sigma^{a}_i)^2$ and   $(\varphi^{a}_i)^2$ mass
terms of the meson Lagrangian, however, not for all contributions.
To ensure that  no terms like $\sigma^a_1\sigma^a_2$ or
 $\varphi^a_1\varphi^a_2$ for strange mesons come from the one-loop quark integrals,
we must impose, in addition to  Eqs.(\ref{condition1}) and (\ref{condition2}), another 
condition 
\be
I_1^{f_{us}}(m_u)+I_1^{f_{us}}(m_s)=0. \label{condition3}
\ee
Conditions (\ref{condition1}), (\ref{condition2}), and (\ref{condition3}) provides 
orthogonality of the ground ($i=1$) and excited ($i=2$) meson states
in the low energy limit $P^2\to 0$ (see Section 2) when $\varphi^a_1$
become Goldstone bosons.

Now let us remind how we fix the basic parameters in the usual
NJL model without excited states of mesons \cite{volk_86}.

To obtain  correct coefficients of kinetic terms
of  mesons in the quark-one-loop approximation, we have to
make the renormalization of
the meson fields 
\be
\sigma_a = g_{\sigma}^a \sigma_a^r, \;\;\;
\varphi_a = g_{\sigma}^a \varphi_a^r, \;\;\;
V^{\mu}_a = \frac{g_V^a}{2} V^{\mu,r}_a, \;\;\; A^{\mu}_a = \frac{g_V^a}{2}
A^{\mu,r}_a,
\label{ren}
\ee
where
\be
g_{\sigma}^{a_{i,j}} = [4 I_2 (m_i, m_j)]^{-1/2}, \;\;\;
I_2 (m_i, m_j) = -i N_c \; \kint \frac{1}{(m_i^2 - k^2)(m_j^2 -
k^2)} ,
\label{g_sigma}
\ee
\be
g_V^a = \sqrt{6} g_{\sigma}^a .
\label{g_V}
\ee
After taking account of the pseudoscalar -- axial--vector
transitions ($\varphi_a \rightarrow A_a$), the additional
renormalization of the pseudoscalar fields
\be
g^a_{\varphi} = Z_a^{-\frac{1}{2}} g_{\sigma}^a,
\label{g_phi}
\ee
appears, where $Z_{\pi} = 1 - 6 m^2_u/M^2_{a_1} \approx 0.7$ for
pions. ($M_{a_1} = 1.23\gev$ is the mass of the axial-vector
$a_1$ meson, \cite{PDG}, $m_u = 280\mev$ (see below and \cite{volk_86})  .
We assume that  $Z_a \approx Z_\pi \approx 0.7$ for any $a$.

After these renormalizations the part of the Lagrangian 
describing the ground states of mesons takes the form
\ba
&&L(\sigma, \phi, V, A)
= - \frac{1}{2}((G^{(-)})^{-1}_{ab} g^{a}_{\sigma}g^{b}_{\sigma} \sigma_a\sigma_b + 
(G^{(+)})^{-1}_{ab}g^{a}_{\varphi}g^{b}_{\varphi}\varphi_a\varphi_b) -
\frac{g_V^{a 2}}{2 G_V} ( V_a^2 + A_a^2 )   \nonumber \\
&&- i N_c~{\rm Tr}~\log \left[ i \partialslash - m +
\left( g^a_\sigma \sigma_a + i \gamma_5 g^a_{\varphi} \varphi_a
+\frac{g_V^a}{2} (\gamma_\mu V^\mu_a +
\gamma_5 \gamma_\mu A^\mu_a ) \right) \tau^a \right].
\label{Lg}
\ea
for simplicity we omitted  the index $r$ of  meson fields.
\par
Lagrangian (\ref{Lg}) in the one-loop approximation results in, the
following expressions for the meson masses
\cite{volk_86}
\ba
M_\pi^2 &=& g_\pi^2 \left[ \frac{1}{G_\pi} - 8 I_1 (m_u) \right] =
\frac{g_\pi^2}{G_\pi} \frac{m_u^0}{m_u}, \;\;\;\;
g_\pi^2 = \frac{1}{4 Z I_2 (m_u, m_u)},
\label{M_pi}
\ea
\be
M_K^2 = g_K^2 \left[ \frac{1}{G_K} - 4 ( I_1 (m_u) + I_1 (m_s) )
\right] + Z^{-1} (m_s - m_u)^2, \nonumber \\
g_K^2 = \frac{1}{4 Z I_2 (m_u, m_s)},\hspace{4cm}
\label{M_K}
\ee
\be
G_\pi=G_{11}^{(+)},\qquad G_K=G_{44}^{(+)},
\ee
\ba
M_{88}^{(-)}&=&g^2_{\eta_u}\left((G^{(+)})^{-1}_{88}-8I_1(m_u)\right),\nonumber\\
M_{99}^{(-)}&=&g^2_{\eta_s}\left((G^{(+)})^{-1}_{99}-8I_1(m_s)\right),\\
M_{89}^{(-)}&=&g_{\eta_u}g_{\eta_s}\left((G^{(+)})^{-1}_{89}\right).
\ea
\be
M_{(\eta,\eta')}^2=\frac12 \left[
M_{88}^{(-)}+M_{99}^{(-)}\mp
\sqrt{(M_{88}^{(-)}-M_{99}^{(-)})^2+4(M_{89}^{(-)})^2}
\right],
\ee
\be
M_\rho^2 = \frac{g_\rho^2}{4 G_V} = \frac{3}{8 G_V
I_2 (m_u, m_u)},\quad
M_\varphi^2 = M_\rho^2 \frac{I_2 (m_u, m_u)}{I_2 (m_s, m_s)},
\label{M_phi}
\ee
\be
M_{K^*}^2 = M_\rho^2 \frac{I_2 (m_u, m_u)}{I_2 (m_u, m_s)} +
\frac{3}{2} (m_s - m_u)^2.
\label{M_K*}
\ee
Now let us fix our basic parameters. For that we  use 
six experimental values \cite{volkov_83,volk_86,volk98}: 
\begin{itemize}
\item[1)] The pion decay constant $F_\pi = 93\mev$ .
\item[2)] The $\rho$-meson decay constant $g_\rho \approx 6.14$. 
Then from the Goldberger-Treiman identity we obtain
\be
m_u = F_\pi g_\pi
\label{GT}
\ee
and from Eqs. (\ref{g_V}) and (\ref{g_phi}) we get
\be
g_\pi = \frac{g_\rho}{\sqrt{6 Z}}, \;\;\;\;
m_u = \frac{F_\pi g_\rho}{\sqrt{6 Z}},\quad m = 280 \mev.
\label{m_u}
\ee
From Eqs. (\ref{g_sigma}) and (\ref{g_V}) we can obtain
(see \cite{Ebert_93})
\be
I_2 (m_u, m_u) = \frac{3}{2 g_\rho^2},\hspace{2cm}
\Lambda_3 = 1.03 \gev.
\label{Lambda_3}
\ee
\item[3)] $M_\pi=135 \mev $, the Eq.(\ref{M_pi}) gives $G_\pi$.
\item[4)] $M_\rho = 770 \mev$, the Eq.(\ref{M_K}) gives $G_V$.
\item[5)] 
\parbox[t]{10cm}{
$
\left.\displaystyle
\begin{array}{l}
M_K \approx 495 \mev,\\ 
M_{\eta'}^2-M_\eta^2
\end{array}\right\}
$ fix $K$ and $m_s$.
}
\end{itemize}

Then the masses of $\eta$, $\eta'$, $K^*$, $\varphi$, 
and scalar mesons
can be calculated with a satisfactory accuracy (see \cite{volk98}).

We can calculate the values of $F_K$ and all the coupling constants 
of strong interactions of  scalar, pseudoscalar, vector, and
axial--vector mesons with each other and with quarks, and
describe  the main decays of these mesons (see \cite{volk_86,volk98}).

Further, when the radial excitations are included, the
parameters will be shifted because of changing the mass formulae.
However, $m_u$ and $\Lambda_3$ will be the same
as they are now. Their numerical values will be calculated in 
Subsec.~3.5.

\subsection{The masses of isovector and strange scalar and pseudoscalar mesons
(ground and excited states)}

After bosonization, the part of  Lagrangian (\ref{L_bos})
describing  the isovector and strange scalar and pseudoscalar mesons takes the form
\ba
&&   {\cal L}(a_{0,1},K_0^*{}_{,1},\pi_1,K_1, a_{0,2},  K_0^*{}_{,2}, \pi_2, K_2)=
     -\frac{a_{0,1}^2}{2G_{a_0}}-\frac{{K_0^*{}_{,1}}^2}{G_{K_0^*}}-\frac{\pi_1^2}{2G_\pi}-
     \frac{K_1^2}{G_K}-\nonumber\\
&&   \frac{1}{2G} ( a_{0,2}^2+2(K_0^*{}_{,2})^2+
     \pi_2^2+2 K_2^2)-\nonumber\\
    &&   i N_c \Tr\ln\left[1+
     \frac{1}{i\partialslash-m}\sum_{a=1}^7\sum_{j=1}^2 \tau_a\left[
     \sigma^a_j+i\gamma_5\varphi^a_j\right]f^a_j
     \right]
\label{bosLag}
\ea
where 
$\sigma^a_j$ and $\varphi^a_j$  are the scalar and pseudoscalar fields:
$\sum_{a=1}^{3}(\sigma^a_j)^2\equiv a_{0,j}^2=({a_{0,j}^0})^2+2a_{0,j}^+a_{0,j}^-$,
$\sum_{a=4}^{7}(\sigma^a_j)^2\equiv 2{\kstar{}_{,j}}^2=
2(\bar{\kstar}_{,j})^0(\kstar{}_{,j})^0+2(\kstar{}_{,j})^+(\kstar{}_{,j})^-$,
$\sum_{a=1}^{3}(\varphi^a_j)^2\equiv \pi_j^2=({\pi_j^0})^2+2\pi_j^+\pi_j^-$,
$\sum_{a=4}^{7}(\varphi^a_j)^2\equiv 2K_j^2=2\bar K^0_j K^0_j+2\bar K^+_jK^-_j$.
As to the coupling constants $G_{aa}$, they
will be defined later (see Subsec.~5 and (\ref{DefG})).

The free part of Lagrangian (\ref{bosLag}) has the following form
\ba
     {\cal L}^{(2)}(\sigma,\varphi)=\frac12\sum_{i,j=1}^2\sum_{a=1}^7
     \left(\sigma^a_i K_{\sigma,ij}^a(P)\sigma^a_j+
     \varphi^a_i K_{\varphi,ij}^a(P)\varphi^a_j\right)
     \label{L2}
\ea
    where the coefficients $K^{a}_{\sigma(\varphi), ij}(P)$ are given by
\ba
    &&  K_{\sigma(\varphi), ij}^{a}(P)=
     -\delta_{ij}\left[\frac{\delta_{i1}}{
G_{aa}^{(\mp)}}+\frac{\delta_{i2}}{G}\right]-\nonumber\\
    &&   i N_c \Tr \int_{\Lambda_3}\frac{d^4k}{(2\pi)^4}
     {1\over \kslash+\Pslash/2-m^a_q}
     r^{\sigma(\varphi)}f_i^a
     {1\over \kslash-\Pslash/2-m^a_{q'}}
      r^{\sigma(\varphi)}f_j^a,
     \label{K_full}
\ea
\be
     r^\sigma=1,\quad r^{\phi}=i\gamma_5,
\ee
\be
     m_q^a = m_u~~(a = 1,...,7);\quad
     m_{q'}^a = m_u~~(a = 1,...,3);~~ m_{q'}^a = m_s~~ (a = 4,...,7),
     \label{m_q^a}
\ee
with $m_u$ and $m_s$ being the constituent quark masses and $f_j^a$ being defined in
(\ref{fDef}).
Integral (\ref{K_full}) is evaluated by expanding in the
meson field momentum $P$. To order $P^2$, one obtains
\ba
     K_{\sigma(\varphi),11}^a(P)&=&
     Z_{\sigma(\varphi),1}^a (P^2 -
     (m_q^a\pm m_{q'}^a)^2- M_{\sigma^a(\varphi^a),1}^2 ),\nonumber\\
     K_{\sigma(\varphi),22}^a(P)
	 &=& Z_{\sigma(\varphi), 2}^a (P^2 -
     (m_q^a\pm m_{q'}^a)^2-  M_{\sigma^a(\varphi^a),2}^2 ),
     \nonumber \\
     K_{\sigma(\varphi),12}^a(P) &=& K_{\sigma(\varphi),21}^a(P) \;\; = \;\;
     \gamma_{\sigma(\varphi)}^a (P^2 - (m^a_q\pm m^a_{q'})^2 )
     \label{Ks_matrix}
\ea
where
\be
     Z_{\sigma,1}^a = 4 I_2^a, \hspace{2em}  Z_{\sigma,2}^a  =  4 I_2^{ff a},
     \hspace{2em} \gamma_{\sigma}^a  = 4 I_2^{f a},
     \label{Zs}
\ee
\be
     Z_{\varphi,1}^a = Z Z_{\sigma,1}^a,
     \hspace{2em}  Z_{\varphi,2}^a  = Z_{\sigma,2}^a  ,
     \hspace{2em} \gamma_{\varphi}^a  = Z^{1/2}\gamma_{\sigma}^a
     \label{Zp}
\ee
and
\ba
      M_{\sigma^a(\varphi^a),1}^2 &=& (Z_{\sigma(\varphi),1}^a)^{-1}
     \left[\frac{1}{G_{aa}^{(\mp)}}-4(I_1(m_q^a) +
     I_1(m_{q'}^a))\right],
     \label{M_1} \\
     M_{\sigma^a(\varphi^a),2}^2 &=& (Z_{\sigma(\varphi),2}^a)^{-1}
     \left[\frac{1}{G}-4(I_1^{ff a}(m_q^a) +
     I_1^{ff a}(m_{q'}^a))\right].
     \label{M_2}
\ea
The factor $Z$ here appears due to
$\pi-a_1$-transitions~\cite{volkov_83,volk_86,volk_97} (see Subsec.~3.1),
and the integrals $I_2^{f..f}$  contain form factors:
\be
     I_2^{f..f_a}(m^a_q,m^a_{q'})={-i N_c\over (2\pi)^4}
     \int_{\Lambda_3} d^4 k
     {f_a({\bf k})..f_a({\bf k})\over ((m^a_q)^2-k^2)((m^a_{q'})^2-k^2)}.
     \label{DefIf}
\ee

After the renormalization of the scalar fields
\be
     \sigma_i^{a r}=\sqrt{Z_{\sigma,i}^a} \sigma_i^{a},\qquad
     \varphi_i^{a r}=\sqrt{Z_{\varphi,i}^a} \varphi_i^{a}  \label{renorm}
\ee
the part of Lagrangian (\ref{L2}) that describes the scalar 
and pseudoscalar mesons
takes the form
\ba
     {\cal L}^{(2)}_{a_0}&=&\frac12
     \left(
     P^2-4 m^2_u -M^2_{a_0, 1}\right)a_{0, 1}^2+ \Gamma_{a_0}\left(
     P^2-4m_u^2\right)a_{0, 1}a_{0, 2}\nonumber\\
     &+&\frac12\left(P^2-4m_u^2- M_{a_{0},2}^2\right)a_{0, 2}^2,
     \label{La0}
\ea
\ba
     {\cal L}^{(2)}_{\kstar}\!&=&\!\frac12\! \left(
     P^2-(m_u+ m_s)^2\! -\! M^2_{\kstar,1}\right)\kstar{}_{,1}^2\!
     +\! \Gamma_{\kstar}
     \left(P^2-(m_u+m_s)^2\right)\kstar{}_{,1}\kstar{}_{,2}\nonumber\\
     &+&\!\frac12\left(P^2-(m_u+m_s)^2- M_{\kstar{},2}^2\right)\kstar{}_{,2}^2,
     \label{LKstar}
\ea
\ba
     {\cal L}^{(2)}_{\pi}&=&\frac12
     \left(
     P^2-M^2_{\pi, 1}\right)\pi_{1}^2+ \Gamma_{\pi}
     P^2\pi_{1}\pi_{2}+
     \frac12\left(P^2- M_{\pi,2}^2\right)\pi_{2}^2,
     \label{Lpi}
\ea
\ba
     {\cal L}^{(2)}_{K}\!&=&\!\frac12\! \left(
     P^2 -\! M^2_{K,1}\right)K_{1}^2\!
     +\! \Gamma_{K}
     P^2 K_{1}K_{2}+
     \!\frac12\left(P^2- M_{K,2}^2\right)K_{2}^2,
     \label{LK}
\ea
where
\be
     \Gamma_{\sigma^a}=\frac{I_2^{f_a}}{\sqrt{I_2 I_2^{ff_a}}},\qquad
     \Gamma_{\varphi^a}=Z^{-1/2}\Gamma_{\sigma^a}.
\ee
After the transformations of the meson fields
\ba
     \sigma^a
     &=& \cos( \theta_{\sigma,a} - \theta_{\sigma,a}^0) \sigma_1^{ar}
     - \cos( \theta_{\sigma,a} + \theta_{\sigma,a}^0) \sigma_2^{ar},   \nonumber \\
     \hat\sigma^{a}
     &=& \sin ( \theta_{\sigma,a} - \theta_{\sigma,a}^0) \sigma_1^{ar}
     - \sin ( \theta_{\sigma,a} + \theta_{\sigma,a}^0) \sigma_2^{ar},
     \label{transf}
\ea
\ba
     \varphi^a
     &=& \cos( \theta_{\varphi,a} - \theta_{\varphi,a}^0) \varphi_1^{ar}
     - \cos( \theta_{\varphi,a} + \theta_{\varphi,a}^0) \varphi_2^{ar},   \nonumber \\
     \hat\varphi^{a}
     &=& \sin ( \theta_{\varphi,a} - \theta_{\varphi,a}^0) \varphi_1^{ar}
     - \sin ( \theta_{\varphi,a} + \theta_{\varphi,a}^0) \varphi_2^{ar}.
     \label{transf1}
\ea
Lagrangians (\ref{La0}), (\ref{LKstar}), (\ref{Lpi}), and (\ref{LK}) 
assume the diagonal form:
\ba
     L_{a_0}^{(2)} &=& \frac12 (P^2 - M_{a_0}^2)~ a_0^2 +
     \frac12 (P^2 - M_{\hat a_0}^2)\hat a_0^{ 2}, \\
     L_{\kstar}^{(2)} &=& \frac12 (P^2 - M_{\kstar}^2)~ \kstar{}^2 +
     \frac12 (P^2 - M_{\hat\kstar}^2)\hat \kstar{}^{ 2}.
     \label{L_pKstar}
\ea
\ba
     L_{\pi}^{(2)} &=& \frac12 (P^2 - M_{\pi}^2)~ \pi^2 +
     \frac12 (P^2 - M_{\hat \pi}^2)\hat \pi^{ 2}, \\
     L_{K}^{(2)} &=& \frac12 (P^2 - M_{K}^2)~ K^2 +
     \frac12 (P^2 - M_{\hat K}^2)\hat K^{ 2}.
     \label{L_pK}
\ea
Here we have
\ba
&&   M^2_{(a_0, \hat a_0)} = \frac{1}{2 (1 - \Gamma^2_{a_0})}
     \biggl[M^2_{a_{0}, 1} + M^2_{a_{0}, 2} \nonumber \\
&&   \qquad \pm\sqrt{(M^2_{a_{0}, 1} - M^2_{a_{0}, 2})^2 +
     (2 M_{a_{0}, 1} M_{a_{0}, 2} \Gamma_{a_0})^2}\biggr]+4m_u^2, \\
&&   M^2_{(\kstar, \hat\kstar)} = \frac{1}{2 (1 - \Gamma^2_{\kstar})}
     \biggl[M^2_{\kstar, 1} + M^2_{\kstar, 2}
       \nonumber \\
&&   \qquad \pm \sqrt{(M^2_{\kstar,1} - M^2_{\kstar,2})^2 +
     (2 M_{\kstar,1} M_{\kstar,2} \Gamma_{\kstar})^2}\biggr]+ (m_u+m_s)^2,
     \label{MpKstar}
\ea
\ba
&&   M^2_{(\pi, \hat \pi)} = \frac{1}{2 (1 - \Gamma^2_{\pi})}
     \biggl[M^2_{\pi, 1} + M^2_{\pi, 2}\pm 
	 \sqrt{(M^2_{\pi, 1} - M^2_{\pi, 2})^2 +
     (2 M_{\pi, 1} M_{\pi, 2} \Gamma_{\pi})^2}\biggr], \\
&&   M^2_{(K, \hat K)} = \frac{1}{2 (1 - \Gamma^2_{K})}
     \biggl[M^2_{K, 1} + M^2_{K, 2}\pm
     \sqrt{(M^2_{K,1} - M^2_{K,2})^2 +
     (2 M_{K,1} M_{K,2} \Gamma_{K})^2}\biggr],
     \label{MpK}
\ea

and

\be
     \tan 2 {\bar{\theta}}_{\sigma(\varphi),a} = \sqrt{\frac{1}{\Gamma_{\sigma^a(\varphi^a)}^2} -
     1}~\left[ \frac{M_{\sigma^a(\varphi^a),1}^2-  
     M_{\sigma^a(\varphi^a),2}^2}{M_{\sigma^a(\varphi^a),1}^2
     + M_{\sigma^a(\varphi^a),2}^2} \right],\qquad
     2 \theta_{\sigma(\varphi),a} = 2 {\bar{\theta}}_{\sigma(\varphi),a} + \pi,
     \label{tan}
\ee
\be
     \sin \theta_{\sigma(\varphi),a}^{0} =\sqrt{{1+\Gamma_{\sigma^a(\varphi^a)}}\over 2}.
     \label{theta0}
\ee
The caret symbol stands for the first radial excitations of mesons.
Transformations (\ref{transf}) and (\ref{transf1}) 
express the ``physical'' fields $\sigma$, $\varphi$,
$\hat\sigma$, and $\hat\varphi$ through the ``bare'' ones $\sigma^{ar}_i$, 
$\varphi^{ar}_i$ and for calculations
these equations must be inverted.
For practical use, we collect
 the values of coefficients in the inverted equations for the scalar and
pseudoscalar fields
 in Table \ref{mixingTable}.
\begin{table}
\caption{The mixing coefficients for the ground and first radially excited states of
scalar and pseudoscalar isovector and strange mesons.
The caret symbol marks the excited states.}
\label{mixingTable}
$$
	\begin{array}{||r|c|c||}
	\hline
		& a_0 & \hat a_0\\
	\hline
	a_{0,1}	&0.87	  &0.82    \\
	a_{0,2} &0.22	  &-1.17   \\
	\hline
	\end{array}
\quad
	\begin{array}{||r|c|c||}
	\hline
		& \kstar & \hat\kstar\\
	\hline
	\kstar{}_{,1}	&0.83	  &0.89    \\
	\kstar{}_{,2} &0.28	  &-1.11   \\
	\hline
	\end{array}
$$
$$
	\begin{array}{||r|c|c||}
	\hline
		& \pi & \hat\pi\\
	\hline
	\pi_1	&1.00	  &0.54    \\
	\pi_2   &0.01	  &-1.14   \\
	\hline
	\end{array}
\quad
	\begin{array}{||r|c|c||}
	\hline
		& K & \hat K   \\
	\hline
	K_1	&0.96	  &0.56   \\
	K_2    & 0.09	  &-1.11  \\
	\hline
	\end{array}
$$

\end{table}


For the weak decay constants of pions and kaons
we obtain 
\ba
F_\pi&=&2m_u\sqrt{ZI_2(m_u)}\cos (\theta_\pi-\theta_\pi^0),\\
F_{\pi'}&=&2m_u\sqrt{ZI_2(m_u)}\sin (\theta_\pi-\theta_\pi^0),
\ea
\ba
F_K&=&(m_u+m_s)\sqrt{ZI_2(m_u,m_s)}\cos (\theta_K-\theta_K^0),\\
F_{K'}&=&(m_u+m_s)\sqrt{ZI_2(m_u,m_s)}\sin (\theta_K-\theta_K^0).
\ea
In the chiral limit we have $\theta_a=\theta_a^0$
and
\ba
&&F_\pi=\frac{m_u}{g_\pi},\quad F_K=\frac{m_u+m_s}{2g_K}\\
&&F_{\pi'}=F_{K'}=0,\quad g_\pi=(Z^\pi_1)^{-1/2},\quad g_K=(Z^K_1)^{-1/2}.
\ea
As one can see from these formulae, in the chiral limit we
obtain the Goldberger-Treiman identities for the coupling 
constants $g_\pi$ and $g_K$. The matrix elements
of  divergences of the axial currents between meson states
and  vacuum equal (PCAC relations) are
\be
\langle 0|\partial^\mu A^a_\mu|\varphi\rangle=M_\varphi^2F_\varphi\delta^{ab},
\ee
\be
\langle 0|\partial^\mu A^a_\mu|\varphi'\rangle=M_{\varphi'}^2F_{\varphi'}\delta^{ab}.
\ee
These axial currents are conserved in the chiral limit because their 
divergences equal zero, according to low-energy theorems.

\subsection{The masses of isoscalar mesons (the ground and excited states)}

The free part of the
effective Lagrangian for isoscalar
scalar and pseudoscalar mesons after bosonization is as follows
\ba
     &&{\cal L}_{\rm isosc}(\sigma,\varphi)=
	-{1\over 2}\sum_{a,b=8}^9\left[
     \sigma^{a}_1
     (G^{(-)})^{-1}_{a b}\sigma^{b}_1 +
     \varphi^{a}_1
     (G^{(+)})^{-1}_{a b}\varphi^{b}_1\right]
\nonumber \\
     &&  -{1\over 2G}\sum_{a=8}^9 \left[\left(\sigma^{a}_2\right)^2 +
     \left(\varphi^{a}_2\right)^2
     \right]
\nonumber \\
     &&-i~{\rm Tr}\ln \left\{1 + {1\over i{ \partialslash} - m}
	\sum_{a=8}^9\sum_{j=1}^2
	\tau^{a}[
     \sigma^{a}_j +
     i\gamma_5
     \varphi^{a}_j
     ]f^{a}_j \right\},  \label{Lbar}
\ea
where $(G^{(\pm)})^{-1}$ is the inverse of $G^{(\pm)}$:
\be
 \begin{array}{ll}
     (G^{(\mp)})^{-1}_{88}=G^{(\mp)}_{88}/D^{(\mp)},\quad &
	(G^{(\mp)})^{-1}_{89}= (G^{(\mp)})^{-1}_{98}=-G^{(\mp)}_{89}/D^{(\mp)}, \\
	(G^{(\mp)})^{-1}_{99}=G^{(\mp)}_{99}/D^{(\mp)},\quad &
	D^{(\mp)}=G^{(\mp)}_{88} G^{(\mp)}_{99}-(G^{(\mp)}_{89})^2 .
 \end{array}
     \label{Tps1}
\ee
From (\ref{Lbar}), in the one-loop approximation, one obtains the
free part of the effective Lagrangian
\ba
     {\cal L}^{(2)}(\sigma,\phi)=\frac12\sum_{i,j=1}^2\sum_{a,b=8}^9
     \left(\sigma^a_i K_{\sigma,ij}^{[a,b]}(P)\sigma^b_j+
     \varphi^a_i K_{\phi,ij}^{[a,b]}(P)\varphi^b_j\right).
	\label{Lisosc}
\ea
The definition of $K_{\sigma(\varphi),i}^{[a,b]}$ is given in
Appendix A.

After  the renormalization of both the scalar and pseudoscalar fields,
 analogous to (\ref{renorm}),  we come to the Lagrangian that
can be represented in a form slightly different from that
of (\ref{Lisosc}). It is convenient to
introduce 4-vectors of ``bare'' fields
\be
     \Sigma=(\sigma_{1}^{8\,r},\sigma_{2}^{8\,r},
	\sigma_{1}^{9\,r},\sigma_{2}^{9\,r}),\qquad
     \Phi=(\varphi_{1}^{8\,r},\varphi_{2}^{8\,r},
	\varphi_{1}^{9\,r},\varphi_{2}^{9\,r}).
\ee
Thus, we have
\ba
     {\cal L}^{(2)}(\Sigma,\Phi)=\frac12\sum_{i,j=1}^4
     \left(\Sigma_i {\cal K}_{\Sigma,ij}(P)\Sigma_j+
     \Phi_i {\cal K}_{\Phi,ij}(P)\Phi_j\right)
     \label{L2a}
\ea
where we introduced new functions ${\cal K}_{\Sigma(\Phi),ij}(P)$ (see Appendix A).
The index $r$ marks  renormalized fields.

Up to this moment we have four pseudoscalar and four scalar meson states
which are the octet and nonet singlets.  Mesons of the same parity
have the same quantum numbers and, therefore,
they are expected to be mixed. In our model the mixing is represented
by $4\times 4$ matrices $R^{\sigma(\varphi)}$ which
transform the ``bare'' fields 
$\sigma_{i}^{8\,r}$,  $\sigma_{i}^{9\,r}$,
$\varphi_{i}^{8\,r}$ and $\varphi_{i}^{9\,r}$
entering  into the 4-vectors  $\Sigma$ and $\Phi$ 
into the ``physical'' ones 
 $\sigma$, $\hat\sigma$, $f_0$, $\hat f_0$ ,
$\eta$,  $\eta'$,  $\hat\eta$, and  $\hat\eta'$
represented as components
of the vectors
 $\Sigma_{\rm ph}$ and $\Phi_{\rm ph}$:
\be
     \Sigma_{\rm ph}=(\sigma,\hat\sigma,f_0,\hat f_0),\qquad
     \Phi_{\rm ph}=(\eta,\hat\eta,\eta',\hat\eta').
\ee
The transformation $R^{\sigma(\varphi)}$ is linear and nonorthogonal:
\be
     \Sigma_{\rm ph}=R^{\sigma}\Sigma,\qquad  \Phi_{\rm ph}=R^{\varphi}\Phi.
\ee
In terms of ``physical'' fields the free part of the effective
Lagrangian is of the conventional form and the coefficients
of matrices $R^{\sigma(\varphi)}$ give the mixing of
the $\bar uu$ and $\bar ss$ components, with and without form factors.

Because of complexity of the procedure of diagonalization for the
matrices of dimensions greater than 2,
there are no such simple formulae as, {\it e.g.},  (\ref{transf}).
Hence, we do not implement it analytically but
use numerical methods to obtain matrix elements
(see Table~\ref{isoscMixTab}).
\begin{table}
\caption{The mixing coefficients for  isoscalar meson states}
\label{isoscMixTab}
$$
	\begin{array}{||r|c|c|c|c||}
	\hline\hline
	 		&\eta 		&\hat\eta 	&\eta' 		&\hat\eta'\\
	\hline
	\varphi^8_{1}	&0.71		&0.62		&-0.32		&0.56		\\
	\varphi^8_{2}   &0.11		&-0.87		&-0.48		&-0.54		\\
	\varphi^9_{1}	&0.62		&0.19		&0.56		&-0.67		\\
	\varphi^9_{2}   &0.06		&-0.66		&0.30		&0.82		\\
	\hline
	\end{array}
$$
$$
	\begin{array}{||r|c|c|c|c||}
	\hline\hline
	 		&\sigma 	&\hat\sigma 	&f_0 		&\hat f_0\\
	\hline
	\sigma^8_{1}	&-0.98		&-0.66		&0.10		&0.17		\\
	\sigma^8_{2}&0.02		&1.15		&0.26		&-0.17		\\
	\sigma^9_{1}	&0.27		&-0.09		&0.82		&0.71		\\
	\sigma^9_{2}&-0.03		&-0.21		&0.22		&-1.08		\\
	\hline
	\end{array}
$$
\end{table}
%

\subsection{The effective Lagrangian for the ground and excited
states of  vector mesons}
The free part of the effective Lagrangian (\ref{L_bos}) describing
the ground and excited states of vector mesons has the form
\ba
L^{(2)} (V) &=&
- \frac12 \sum_{i, j = 1}^{2} \sum_{a = 1}^{9} V_i^{a \mu} (P)
R_{ij}^{a \mu\nu} (P) V_j^{a \nu}(P) ,
\label{LV_2}
\ea
where
\ba
&&\sum_{a = 1}^{3} (V_i^{a \mu })^2 = 
(\rho_i^{0 \mu})^2 +
2 \rho_i^{+ \mu} \rho_i^{- \mu},~~~
(V_i^{4 \mu})^2 + (V_i^{5 \mu})^2 = 2 K_i^{* + \mu} K_i^{* - \mu}, \nonumber \\
&&(V_i^{6 \mu})^2 + (V_i^{7 \mu})^2 = 2 K_i^{* 0 \mu}
K_i^{* 0 \mu},~~~ (V_i^{8 \mu})^2 = (\omega_i^\mu)^2,\quad
(V_i^{9 \mu})^2 = (\varphi_i^\mu)^2
\label{V^a}
\ea
and
\ba
R_{ij}^{a \mu \nu} (P) =
-~ \frac{\delta_{ij}}{G_V} g^{\mu\nu}\hspace{5cm}
\nonumber \\
-~ i~ N_{\rm c} \; {\rm tr}\, \kint \left[
\frac{1}{\kslash + \half\Pslash - m_q^a}\gamma^\mu f_i^{a,V}
\frac{1}{\kslash - \half\Pslash - m_{q'}^a}  \gamma^\nu f_j^{a,V}
\right]  , 
\label{R_full}
\ea
To order $P^2$, one obtains
\ba
R_{11}^{a \mu\nu } &=& W_1^a [P^2 g^{\mu\nu} - P^\mu P^\nu -
g^{\mu\nu} (\bar M^a_1)^2], \nonumber \\
R_{22}^{a\mu\nu} &=& W_2^a [P^2 g^{\mu\nu} - P^\mu P^\nu -
g^{\mu\nu} (\bar M^a_2)^2], \nonumber \\
R_{12}^{a\mu\nu} &=& R_{21}^{\mu\nu a} = \bar\gamma^a
[P^2 g^{\mu\nu} - P^\mu P^\nu - \frac{3}{2} (m_q^a-m_{q'}^a)^2 g^{\mu\nu}].
\label{R_ij}
\ea
Here
\ba
W_1^a &=& \frac{8}{3} I_2^a,~~~W_2^a = \frac{8}{3} I_2^{ff a},~~~
\bar\gamma^a = \frac{8}{3} I_2^{f a}, \label{Wdef}\\
(\bar M_1^a)^2 &=& (W_1^a G_V)^{-1} + \frac{3}{2}
(m_q^a-m_{q'}^a)^2, \\
(\bar M_2^a)^2 &=& (W_2^a G_V)^{-1} + \frac{3}{2}
(m_q^a-m_{q'}^a)^2.
\label{WM}
\ea
After renormalization of the meson fields
\be
V_i^{a r \mu} = \sqrt{W_i^a}~V_i^{a \mu}
\label{V^r}
\ee
we obtain the Lagrangians
\ba
L_\rho^{(2)} &=& - \frac12 \left[\left( g^{\mu\nu} P^2 - P^\mu P^\nu -
g^{\mu\nu} M^2_{\rho_1}\right) \rho^\mu_1 \rho^\nu_1\right. \nonumber \\
&+& \left.2 \Gamma_\rho  \left( g^{\mu\nu} P^2 - P^\mu P^\nu\right) \rho_1^\mu
\rho_2^\nu + \left( g^{\mu\nu} P^2 - P^\mu P^\nu -
g^{\mu\nu} M^2_{\rho_2}\right) \rho^\mu_2 \rho^\nu_2 \right],
\label{L2_V1}
\ea
\ba
L_\varphi^{(2)} &=& - \frac12 \left[\left( g^{\mu\nu} P^2 - P^\mu P^\nu -
g^{\mu\nu} M^2_{\varphi_1}\right) \varphi^\mu_1 \varphi^\nu_1 \right.\nonumber \\
&+&\left. 2 \Gamma_\varphi  \left( g^{\mu\nu} P^2 - P^\mu P^\nu\right) \varphi_1^\mu
\varphi_2^\nu + \left( g^{\mu\nu} P^2 - P^\mu P^\nu -
g^{\mu\nu} M^2_{\varphi_2}\right) \varphi^\mu_2 \varphi^\nu_2 \right],
\label{L2_V2}
\ea
\ba
L_{K^*}^{(2)} &=& - \frac12 \left[\left( g^{\mu\nu} P^2 - P^\mu P^\nu -
g^{\mu\nu} \left(\frac{3}{2} (m_q^a-m_{q'}^a)^2 + M^2_{K^*_1}\right)\right) K^{*\mu}_1
K^{*\nu}_1 \right.\nonumber \\
&+&  2 \Gamma_{K^*} \left( g^{\mu\nu} P^2 - P^\mu P^\nu -
g^{\mu\nu} \frac{3}{2}
(m_q^a-m_{q'}^a)^2\right) K_1^{*\mu} K_2^{*\nu} \nonumber \\
&+&\left. \left( g^{\mu\nu} P^2 - P^\mu P^\nu -
g^{\mu\nu} \left(\frac{3}{2} (m_q^a-m_{q'}^a)^2 + M^2_{K^*_2}\right)\right) 
K^{*\mu}_2 K^{*\nu}_2 \right].
\label{L2_V3}
\ea
Here
\ba
M_{\rho_1}^2 = \frac{3}{8 G_V I_2(m_u, m_u)},~~~
M_{{K^*}_1}^2 = \frac{3}{8 G_V I_2(m_u, m_s)},  \nonumber \\
M_{\varphi_1}^2 = \frac{3}{8 G_V I_2(m_s, m_s)},~~~
M_{\rho_2}^2 = \frac{3}{8 G_V I^{ff}_2(m_u, m_u)}, \nonumber \\
M_{{K^*}_2}^2 = \frac{3}{8 G_V I^{ff}_2(m_u, m_s)},~~~
M_{\phi_2}^2 = \frac{3}{8 G_V I^{ff}_2(m_s, m_s)},
\label{MV_i}
\ea
\be
\Gamma_{a_{i,j}} = \frac{I_2^{f a}(m_i, m_j)}
{\sqrt{I_2^a(m_i, m_j)I_2^{ff a}(m_i, m_j)}}.
\label{GammaV}
\ee
After transformations of the vector meson fields, similar to Eqs.
(\ref{transf}) for the pseudoscalar mesons,  Lagrangians
(\ref{L2_V1}), (\ref{L2_V2}), (\ref{L2_V3}) take the diagonal form
\ba
L^{(2)}_{V^a, \bar V^a} = - \frac12 \left[ (g^{\mu\nu} P^2 -
P^\mu P^\nu - M^2_{V^a} ) V^{a \mu} V^{a \nu}  \right. \nonumber \\
\left. + (g^{\mu\nu} P^2 - P^\mu P^\nu -
M^2_{\bar V^a} ) \bar V^{a \mu} \bar V^{a \nu} \right],
\label{LDV}
\ea
where $V^{a\mu}$ and $\bar V^{a\mu}$ are  physical ground and excited
states of vector mesons
\ba
M^2_{\rho, \bar\rho} &=& \frac{1}{2 (1 - \Gamma^2_\rho)}~
\left[M^2_{\rho_1} + M^2_{\rho_2}~ \mp~ \sqrt{(M^2_{\rho_1} -
M^2_{\rho_2})^2 + (2 M_{\rho_1}M_{\rho_2} \Gamma_\rho)^2}
\right]  \nonumber \\
&=& M^2_{\omega, \bar\omega},
\label{Mrho}
\ea
\be
M^2_{\varphi, \bar\varphi} = \frac{1}{2 (1 - \Gamma^2_\phi)} \left[
M^2_{\varphi_1} + M^2_{\phi_2} \mp \sqrt{(M^2_{\varphi_1} -
M^2_{\varphi_2})^2 + (2 M_{\varphi_1}M_{\varphi_2}
\Gamma_\varphi)^2} \right] ,
\label{Mphi}
\ee
\ba
M^2_{K^*, \bar K^*} = \frac{1}{2 (1 - \Gamma^2_{K^*})} \left[
M^2_{K^*_1} + M^2_{K^*_2} + 3 \Delta^2 (1 - \Gamma^2_{K^*})
\right.\nonumber \\
\left.
\mp \sqrt{(M^2_{K^*_1} - M^2_{K^*_2})^2 +
(2 M_{K^*_1}M_{K^*_2} \Gamma_{K^*})^2} \right].
\label{MK^*}
\ea

\subsection{Numerical estimates.}

In our model we have six basic parameters (see Subsection~3.1): the masses of
the constituent $u(d)$ and $s$ quarks, $m_u=m_d$ and
$m_s$, the cut-off parameter $\Lambda_3$, two four-quark
coupling constants (one for the scalar and pseudoscalar 
channels, $G$, and the other for the vector and 
axial-vector channels, $G_V$)  and the  't Hooft coupling constant
$K$. We  fixed these parameters with the help  of
input parameters: the pion decay constant $F_\pi=93 \mev$,
the $\rho$-meson decay constant $g_\rho=6.14$
(decay $\rho\to2\pi$)%
\footnote{Here,  we  used  the relation $g_\rho=\sqrt{6}g_{\sigma}$
 together with the Goldberger--Treiman relation
$g_\pi=m/F_\pi=Z^{-1/2}g_{\sigma}$ to fix the parameters
$m_u$ and $\Lambda_3$.
},
the masses of pion, kaon, $\rho$-meson, and the mass difference of $\eta$ and $\eta'$
mesons. Using mass formulae given in previous subsections of this Section,
we obtain numerical estimates of these parameters:
\ba
	&&m_u=280\mev,\quad m_s=405 \mev, \quad\Lambda_3=1.03 \gev, \nonumber\\
	&&G=3.14\gev^{-2}, \quad G_V=12\gev^{-2}, \quad K=6.1\gev^{-5}.
\ea
When excited meson states are introduced, a set of additional 
parameters related to the form factors  appears in our model:
the slope parameters $d_{qq}$ and the external parameters $c^U_{qq}$.
The slope parameters $d_{qq}$  
are fixed by special conditions (see Eqs.(\ref{condition1}),
(\ref{condition2}), (\ref{condition3}))
from which we obtain: 
$d_{uu}=-1.78 \gev^{-2}$,
$d_{us}=-1.76 \gev^{-2}$,  $d_{ss}=-1.73 \gev^{-2}$.
As it was mentioned earlier, we assume here that $d_{uu}$, $d_{us}$,
and $d_{ss}$ do not depend on parity and spin of mesons.    

The parameters $c^{\sigma^a(\varphi^a)}_{qq}$ are fitted by masses of
excited pseudoscalar and vector mesons,
 $c_{uu}^{\pi,a_0}=1.44$, $c_{uu}^{\eta,\eta',\sigma,f_0}=1.5$,
$c_{us}^{K,\kstar}=1.59$,
$c_{ss}^{\eta,\eta',\sigma,f_0}=1.66$, $c_{uu}^{\rho}=1.33$,
$c_{us}^{K^*}=1.6$, $c_{ss}^{\varphi}=1.41$.
These parameters characterize how stronger 
the quark currents with form factors attract to each other  than 
those without form factors.
We use the same parameters for the scalar and pseudoscalar mesons 
(global chiral symmetry). This allows us to
 predict  the masses of  ground and excited states of scalar mesons. 
The result is represented in Table \ref{masses} together with experimental values.
\begin{table}
\caption{The model masses of mesons, MeV}
\label{masses}
$$
\begin{array}{||l|c|c|c|c||}
	\hline \hline
		& GR 		& EXC		& GR(Exp.)\,\cite{PDG}		&EXC(Exp.)\,
\cite{PDG} 	\\
	\hline
M_{\sigma} 	& 530		& 1330		&400-1200		&1200-1500	\\
M_{f_0}		& 1070		& 1600		&980\pm10		&1712\pm5	\\
M_{a_0}		& 830		& 1500		&983.4\pm0.9		&1474\pm19	\\
M_{\kstar} 	& 960		& 1500		&905\pm50\,\cite{ishida}&1429\pm12	\\
M_{\pi} 	& 140		& 1300		&139.56995\pm0.00035	&1300\pm100	\\
M_{K} 		& 490		& 1300		&497.672\pm0.031	&1460(?)	\\
M_{\eta} 	& 520		& 1280		&547.30\pm0.12		&1297.8\pm2.8	\\
M_{\eta'} 	& 910		& 1470		&957.78\pm0.14		&1440-1470	\\
M_{\rho}	& 770		& 1470		&770.0\pm0.8		&1465\pm25	\\
M_{\varphi}	& 1019		& 1682		&1019.413\pm0.008	&1680\pm20	\\
M_{K^*}		& 887		& 1479		&891.59\pm0.24		&1412\pm12	\\
	\hline \hline
\end{array}
$$
\end{table}

We also calculate
the angles $\theta_a$
and $\theta_a^0$:
\ba
\begin{array}{lll}
\theta_{\pi}=59.48^{\circ} &
\theta_{\pi}^0=59.12^{\circ}, &
\theta_{K}=60.2^{\circ},\\
\theta_{K}^0=57.13^{\circ},&
\theta_{\rho}=81.8^{\circ},&
\theta_{\rho}^0=81.5^{\circ}\\
\theta_{K^{*}}=84.7^{\circ},&
\theta_{K^{*}}^0=59.14^{\circ},&
\theta_{\varphi}=68.4^{\circ},\\
\theta_{\varphi}^{0}=57.13^{\circ},&
\theta_{a_0}=72.0^{\circ}, &
\theta_{a_0}^0=61.5^{\circ}, \\
\theta_{\kstar}=74.0^{\circ}&
\theta_{\kstar}^0=60.0^{\circ}.
\end{array}
\label{angles}
\ea
We consider it expedient to give the values of  angles
because they will be used in the next Section when the calculation of
strong  decays of the ground and first radially excited states of the $\pi$ and
$\rho$ meson will be treated in detail. However,
 the mixing coefficients for $\pi$, $K$, $a_0$,
and $\kstar$ defined by these angles have
 been displayed in Table~\ref{mixingTable}.
The mixing coefficients for $\eta$, $\eta'$, $\sigma$, and
$f_0$ are given in Table~\ref{isoscMixTab}.

Having fixed all parameters in our model, 
we can predict the masses of  $\eta$, $\eta'$, $\kstar$, and $\varphi$
mesons and  all masses of the ground and first radially excited scalar
meson states. We also calculate the weak decay constants
for the pion and kaon (both for the ground and excited states):
\be
F_\pi = 93 \mev,\quad  F_{\pi'} = 0.57~ \mev,
\label{ff'}
\ee
\be
F_K = 1.16 F_\pi = 108 \mev, \quad F_{K'} = 3.3~ \mev.
\label{fKK'}
\ee
Moreover, now we are able to estimate all strong coupling constants
for the mesons considered in this paper. In the next Section we
calculate some of these constants that define the strong decay processes
of  ground states and first radial excitations of the
 scalar, pseudoscalar, and vector meson nonets.

\section{Strong decays of mesons.}

\subsection{Decays $\rho \to 2 \pi, \pi' \to \rho \pi, \pi'\to\sigma\pi, \rho' \to 2\pi,
\rho' \to \omega \pi$ and $\omega' \to \rho \pi$.}
\par
In this section we  calculate the  widths of 
main decays of scalar, pseudoscalar, and vector meson nonets
(for Subsec. 4.1 see refs.\cite{ven},
for Subsec. 4.2  ref.\cite{strange}, for Subsecs. 4.3 and 4.4 see ref.~\cite{volk_99B}) 
through triangle quark diagrams.
When calculating these diagrams, we keep the least possible
dependence on  external momenta: squared for the anomaly type
graphs and linear for  nother types. We omit  the higher order
momentum dependence.

We start with the
decay $\rho \to 2\pi$. The amplitude describing this decay has
the form
\be
T_{\rho \to 2\pi} = i~\frac{g_\rho}{2}~\epsilon_{ijk}~(p_j -
p_k)^\nu~\rho^i_\nu \pi^j \pi^k,
\ee
where $p_{j,k}$ are  pion momenta and $\epsilon_{ijk}$ is 
antisymmetric tensor. Using the value $\alpha_{\rho} =
{g^2_{\rho}}/(4 \pi) \approx 3~~(g_\rho \approx 6.1)$ of refs.
\cite{volkov_83,volk_84,volk_86,ebert_86} we obtain for the decay width
\be
\Gamma_{\rho \to 2\pi} =  \frac{\alpha_\rho}{12~M_\rho^2}~
(M_\rho^2 - 4~M_\pi^2)^{3/2} \approx 151.5\mev.
\ee
The experimental value is \cite{PDG}
\be
\Gamma_{\rho \to 2\pi} = 150.7 \pm 1.2~{\rm MeV}.
\ee
\par
Now let us calculate this amplitude in our model with the excited
states of mesons. To this end, we rewrite the amplitude $T_{\rho \to
2\pi}$ in the form
\be
T_{\rho \to 2\pi} = i~c_{\rho \to 2\pi}~\epsilon_{ijk}~(p_j -
p_k)^\nu~\rho^i_\nu \pi^j \pi^k,
\ee
and calculate the factor $c_{\rho \to 2\pi}$ in the new model.
Using Eqs. (\ref{renorm}), (\ref{transf1}) and (\ref{V^r}) we
can find the following expressions for  meson fields $\pi_i$
and $\rho_i$ from the Lagrangian (\ref{L_bos})  expressed in terms of the
physical states $\pi, \pi'$ and $\rho, \rho'$ 
\ba
\pi_1 &=&\frac{ \sin(\theta_{\pi}+\theta_{\pi}^0) \pi - \cos(\theta_{\pi}+\theta_{\pi}^0)
\pi'}{\sqrt{Z_1} \sin2\theta_{\pi}^0}, \nonumber \\
\pi_2 &=&\frac{ \sin(\theta_{\pi}-\theta_{\pi}^0) \pi - \cos(\theta_{\pi}-\theta_{\pi}^0)
\pi'}{\sqrt{Z_2} \sin2\theta_{\pi}^0},
\label{pi}
\ea
\ba
\rho_1& =& \frac{\sin(\theta_{\rho}+\theta_{\rho}^0) \rho - \cos(\theta_{\rho}+\theta_{\rho}^0)\rho'}
{\sin2\theta_{\rho}^0 \sqrt{8/3~I_2}}, \nonumber \\
\rho_2& =& \frac{\sin(\theta_{\rho}-\theta_{\rho}^0) \rho - \cos(\theta_{\rho}-\theta_{\rho}^0)\rho'}
{\sin2\theta_{\rho}^0 \sqrt{8/3~I_{2,\rho}^{ff}}},
\label{ro}
\ea
or, using the values $I_2 = 0.04, I_{2,\rho}^{ff} = 0.0244$,
and $\theta_\pi$, $\theta_\pi^0$, $\theta_\rho$ and $\theta_\rho^0$
from Eqs.~(\ref{angles}),
we obtain \footnote{Analogous formulae are obtained for the $\omega$-meson.}
\ba
\pi_1 = \frac{0.878 \pi +0.48 \pi'}{0.88 \sqrt{Z_1}},~~~
\pi_2 = \frac{0.0061 \pi - \pi'}{0.88 \sqrt{Z_2}}, \nonumber \\
\rho_1 = (0.744 \rho + 0.931 \rho')~g_\rho/2,~~~
\rho_2 = (0.48~ \rho - 1.445~\rho')~g_\rho/2.
\label{piro}
\ea
The decay $\rho \to 2 \pi$ is described by the quark triangle
diagrams with the vertices \\
$\rho_1 (\pi^2_1 + 2\pi_1\pi_2 + \pi_2^2)$ and
$\rho_2 (\pi^2_1 + 2\pi_1\pi_2 + \pi_2^2)$ (see Fig.~\ref{fig:graph}).
Using Eqs. (\ref{pi}), (\ref{ro}) and (\ref{piro}), we arrive at the factor
\footnote{Taking account of the $\pi \to a_1$ transitions on
 external pion lines, we obtain additional factors $Z$
($\bar{Z}$) in the numerators of our triangle diagrams which
cancel corresponding factors in $Z_i$ (see Eqs. (\ref{I_12}),
(\ref{pi}) and ref. \cite{volk_86}). Therefore, in future
we shall ignore the factors $Z$ ($\bar{Z}$) in $Z_i$. }
$c_{\rho \to 2\pi}$ 
\ba
c_{\rho \to 2\pi} = c_{\rho_1 \to 2\pi} + c_{\rho_2 \to 2\pi} =
0.975~g_\rho/2,
\ea
\ba
c_{\rho_1 \to 2\pi} &=& \frac{\sin(\theta_{\rho} + \theta_{\rho}^0)}{\sin^2
2\theta_{\pi}^0~\sin 2\theta_{\rho}^0~\sqrt{8/3~I_2}}~[(\sin(\theta_{\pi} +
\theta_{\pi}^0))^2 + 2 \sin(\theta_{\pi} + \theta_{\pi}^0) \sin(\theta_{\pi} - \theta_{\pi}^0)
\Gamma_\pi   \nonumber  \\
&+& (\sin(\theta_{\pi} - \theta_{\pi}^0))^2 = \sin^2 2\theta_{\pi}^0] =
\frac{\sin(\theta_{\rho} + \theta_{\rho}^0)}{\sin 2\theta_{\rho}^0~\sqrt{8/3~I_2}}
= 0.745~g_\rho/2,  \nonumber \\
c_{\rho_2 \to 2\pi} &=& \frac{sin(\theta_{\rho} - \theta_{\rho}^0)}{\sin^2
2\theta_{\pi}^0~ \sin 2\theta_{\rho}^0~\sqrt{8/3~I_{2,\rho}^{ff}}}~
[(\sin(\theta_{\pi} + \theta_{\pi}^0))^2~\frac{I_2^f}{I_2}  \nonumber \\
&+& 2 \sin(\theta_{\pi} + \theta_{\pi}^0) \sin(\theta_{\pi} -
\theta_{\pi}^0) \frac{I_2^{ff}}{\sqrt{I_2~I_2^{ff}}} +
(\sin(\theta_{\pi} - \theta_{\pi}^0))^2 \frac{I_2^{fff}}{I_2^{ff}}] = 0.227~
g_\rho/2.
\label{cro}
\ea
Here we used the values
$I_2^f = 0.0185,~I_2^{ff} = 0.0289,~I_2^{fff} =
0.0224$ and the equation $\Gamma_\pi = - \cos 2\theta_{\pi}^0$ (
it can  easily be derived from Eq.~(\ref{theta0})). 
Then the decay width $\rho \to 2 \pi$ is equal to
\be
\Gamma_{\rho \to 2\pi} \approx 149~{\rm MeV}.
\ee
In the limit $f = 0$ ($\theta_{\pi} = \theta_{\pi}^0, \theta_{\rho} = \theta_{\rho}^0$) from
Eqs. (\ref{cro}) one finds
\be
c_{\rho \to 2\pi} = c_{\rho_1 \to 2\pi} = g_\rho/2,~~~
c_{\rho_2 \to 2\pi} = 0.
\ee
\par
Now let us consider the decay $\pi' \to \rho \pi$. The amplitude
of this decay is of the form
\be
T_{\pi' \to \rho \pi}^{\nu} = i~c_{\pi' \to \rho \pi}~\epsilon_{ijk}~
(p_j + p_k)^\nu~\rho^i_\nu \pi^j \pi^k,
\ee
where
\be
c_{\pi' \to \rho \pi} = c_{\pi' \to \rho_1 \pi} + c_{\pi' \to
\rho_2 \pi}.
\ee
Then for $c_{\pi' \to \rho_1 \pi}$ we obtain
\ba
c_{\pi' \to \rho_1 \pi} &=& \frac{2}{(\sin 2\theta_{\pi}^0)^2}~
[-\sin(\theta_{\pi}+\theta_{\pi}^0) \cos(\theta_{\pi}+\theta_{\pi}^0) - \sin 2\theta_{\pi}~
\Gamma_\pi - \sin(\theta_{\pi}-\theta_{\pi}^0)
\cos(\theta_{\pi}-\theta_{\pi}^0)  \nonumber  \\
&=& - \sin 2\theta_{\pi} \cos 2\theta_{\pi}^0 + \sin 2\theta_{\pi} \cos 2\theta_{\pi}^0 = 0]~
\frac{\sin(\theta_{\rho}+\theta_{\rho}^0)}{\sin 2\theta_{\rho}^0}~g_\rho/2 = 0,
\label{cro1}
\ea
\ba
c_{\pi' \to \rho_2 \pi} = \frac{2}{(\sin 2\theta_{\pi}^0)^2}~
[-\sin(\theta_{\pi}+\theta_{\pi}^0) \cos(\theta_{\pi}+\theta_{\pi}^0) \frac{I_2^f}{I_2} -
\sin 2\theta_{\pi} \frac{I_2^{ff}}{\sqrt{I_2~I_2^{ff}}} \nonumber \\
- \sin(\theta_{\pi}-\theta_{\pi}^0) \cos(\theta_{\pi}-\theta_{\pi}^0)
\frac{I_2^{fff}}{I_2^{ff}}]~
\frac{\sin(\theta_{\rho} - \theta_{\rho}^0)}{\sin 2\theta_{\rho}^0}~
\sqrt{\frac{I_2}{I_2^{ff}}}~g_\rho/2 = - 0.573~g_\rho/2.
\label{cro2}
\ea
For the decay width $\pi' \to \rho \pi$ we get
\ba
\Gamma_{\pi' \rightarrow \rho \pi} &=& \frac{c_{\pi' \to \rho
\pi}^2}{4\pi M^3_{\pi'}M^2_{\rho}}\Phi(M_{\pi'},M_{\rho},M_{\pi})^3 \approx 220~{\rm MeV}
\label{Gpi'1}
\ea
where
\be
\Phi(M_1,M_2,M_3)=\sqrt{M^4_{1} + M^4_{2}
+ M^4_{3} - 2(M^2_{1}M^2_{2} + M^2_{1}M^2_{3} +
M^2_{2}M^2_{3} )}.\label{PhiMMM}
\ee
The decay $\pi' \to \sigma \pi$ is calculated in a similar way as 
$\hat\eta\to a_0\pi$ (see Subsec. 4.4). Here, we need the mixing coefficients for 
the scalar meson given in Table~\ref{isoscMixTab}.
We omit details and obtain
\be
\Gamma_{\pi'\to\sigma\pi}\approx 80 \mev,
\ee
therefore, the total width is estimated as
\be
\Gamma^{\rm tot}_{\pi}\approx300 \mev,
\ee
This value is in agreement with the
experimental data \cite{PDG}
\be
\Gamma^{\rm tot}_{\pi'} = 200 - 600~ {\rm MeV}.
\ee
\par

 For the decay $\rho' \to 2\pi$
we arrive in our model at the result
\be
\Gamma_{\rho' \to 2\pi} \approx 22~{\rm MeV}.
\ee
Most of our results are in agreement with  results of the relativized
potential quark model with the $3P_0$-mechanism of meson decays \cite{geras}.
\par
 To conclude this subsection, we calculate the decay widths of 
processes $\rho' \to \omega \pi$ and $\omega' \to \rho \pi$.
These decays go through anomalous triangle quark loop diagrams.
The amplitude of the decay $\rho' \to \omega \pi$ takes the form
\be
T^{\mu \nu}_{\rho' \to \omega \pi} = \frac{3 \alpha_{\rho}
c_{\rho' \to \omega \pi}}{2 \pi F_{\pi}}~\epsilon^{\mu \nu \rho \sigma}~
q_{\rho} p_{\sigma},
\ee
where $q$ and $p$ are  momenta of the $\omega$ and $\rho'$ meson,
respectively. The factor $c_{\rho' \to \omega \pi}$ is similar to the
factors $c_{\rho \to 2\pi}$ and $c_{\pi' \to \rho \pi}$ in the previous
equations and arises from the four triangle quark diagrams with vertices
$\pi_1(\rho_1 \omega_1 + \rho_2 \omega_1 + \rho_1 \omega_2 +
\rho_2 \omega_2)$
\footnote{We  neglect the diagrams with vertices $\pi_2$, because their
contribution to the ground state of the pion is very small (see
Eq.(\ref{piro})).}. Using the estimate
\be
c_{\rho' \to \omega \pi} \approx - 0.3,
\ee
we obtain for the decay width
\ba
\Gamma_{\rho' \rightarrow \omega \pi} &=& \frac{3}{2 \pi M^3_{\rho'}}~
\left(\frac{\alpha_{\rho}~c_{\rho' \to \omega \pi}}{8~\pi~F_{\pi} }\right)^2~
\Phi(M_{\rho'},M_{\omega}, M_{\pi})^3\approx 75\mev.
\ea
For the decay $\omega' \to \rho \pi$ we have the relation
\be
\Gamma_{\omega' \to \rho \pi} \approx 3~\Gamma_{\rho' \to \omega \pi}
\ee
leading to the estimate
\be
\Gamma_{\omega' \to \rho \pi} \approx 225~{\rm MeV}.
\ee
The experimental values are \cite{clegg}
\be
\Gamma^{exp}_{\rho' \to \omega \pi} =
0.21~\Gamma^{\rm tot}_{\rho'} = 65.1~\pm~12.6~{\rm MeV}
\ee
and \cite{PDG}
\be
\Gamma^{exp}_{\omega' \to \rho \pi} = 174~\pm~60~{\rm MeV}.
\ee
Finally, let us quote the ratio of the decay widths $\rho' \to \omega \pi$
and $\rho' \to 2\pi$
\be
\frac{\Gamma_{\rho' \to 2 \pi}}{\Gamma_{\rho' \to \omega \pi} } \approx 0.3,
\ee
which is to be compared with the experimental value 0.32 (see \cite{clegg}).
\par
Thus, we can see that all our estimates are in satisfactory agreement with
experimental data.
\par

 Our calculations have shown that the main decay of the $\rho$-meson,
$\rho \to 2\pi$, changes very little after including the excited meson
states into the NJL model. The main part of this decay (75\%) comes from
the $\rho$-vertex without the form factor, whereas the remaining 25\% of the
decay are due to the $\rho$-vertex with the form factor. As a result,  the new
coupling constant $g_{\rho}$ turns out to be very close to the former value.
\par
For the decay $\pi' \to \rho \pi$ we meet an  opposite situation.
Here the channel connected with the $\rho$-vertex without the form factor
is closed because the states $\pi$ and $\pi'$ are orthogonal to
each other, and the total decay width of
$\pi' \to \rho \pi$ is defined by the channel going through the $\rho$-
vertex with the form factor. As a result, we obtain
the quoted  value that
satisfies  experimental data \cite{PDG}.
The decay $\pi' \to \sigma \pi$ gives a noticeble correction to
the total decay width of $\pi'$.  These results disagee 
with the results obtained in the relativized version of the $3P_1$ potential
model \cite{geras} in the subject of the $\pi'\to\sigma\pi$ decay mode.

For the decay $\rho' \to 2\pi$ we obtain  strong compensation
of the contributions from the two channels, related to $\rho$-vertices
with and without form factors, and the corresponding  decay width is equal to
22 MeV. This value is very close to the result of ref.\cite{geras}.

It should be emphasized that the decays $\rho' \to \omega \pi$
and $\omega' \to
\rho \pi$ belonging to a different class of quark loop diagrams
(``anomaly diagrams'')
are also satisfactorily described by our model.

\subsection{The decays of strange mesons (vectors and pseudoscalars).}

In the framework of our model, the  decay modes
of  excited  mesons 
are represented by triangle diagrams with form factors.
A total set of diagrams similar to those in 
Fig.~\ref{fig:graph} can be represented as one graph:
a triangle with shaded angles (see Fig.~\ref{trigX}).  
Every vertex in such diagrams is  momentum-dependent and includes
form factors defined  in Subsection~3.1.
For the strange vector and pseudoscalar mesons being decaying,  
each black shaded vertex with a pseudoscalar meson
is implied to contain
the following linear combination for the ground state:
	\be
	\bar f_a={1\over \sin 2\theta_{a}^{0}} \left[
	{\sin(\theta_a+\theta_{a}^{0}) \over \sqrt{Z_1^a}}+
	{\sin(\theta_a-\theta_{a}^{0}) \over \sqrt{Z_2^a}}f_a
	\right], \label{physformfactors1}
	\ee
and for an excited state,
	\be
	\bar f_{a}^{\prime}={-1\over \sin 2\theta_{a}^{0}} \left[
	{\cos(\theta_a+\theta_{a}^{0}) \over \sqrt{Z_1^a}}+
	{\cos(\theta_a-\theta_{a}^{0}) \over \sqrt{Z_2^a}}f_a
	\right], \label{physformfactors2}
	\ee
where
$\theta_a$ and $\theta_{a}^{0}$ are the angles defined in subsection~3.5
(see Eqs.~(\ref{tan}), (\ref{theta0}), and (\ref{angles})) and
$f_a$ is one of the form factors
defined in subsection~3.1 (see Eq. (\ref{fDef})).
For vector meson vertices, we have the same linear combinations
except that $Z_i^a$ are to be replaced by $W_i^a$ (\ref{Wdef}), and
the related angles and form factor parameters must be chosen.

Now we can calculate the decay widths of 
excited mesons. Let us start with the process
$K^{*'}\to K^*\pi$.
The corresponding amplitude, $T_{K^{*'}\to K^{*}\pi}^{\mu\nu}$, has the form
	\be
	T_{K^{*'}\to K^{*}\pi}^{\mu\nu}= g_{K^{*'}\to K^*\pi}
	\epsilon^{\mu\nu\alpha\beta}p_{\alpha}q_{\beta}
	\label{T_Kstar_to_Kstarpi}
	\ee
where $p$  and $q$ are  momenta of the $K^{*'}$- and $K^{*}$-mesons,
respectively,
and $g_{K^{*'}\to K^*\pi}$  is the (dimensional) coupling
constant that follows from the combination of one-loop integrals
	\be
	g_{K^{*'}\to K^*\pi}=
	{8 m_s\over m_u^2-m_s^2}
	\left(
	\J_{2,0}[\bar f_{K^{*}}^\prime\bar f_{K^*}\bar f_{\pi}]-
	\J_{1,1}[\bar f_{K^{*}}^\prime\bar f_{K^*}\bar f_{\pi}]
	\right).
	\label{KstartoKstarpi}
	\ee

In Eq.(\ref{KstartoKstarpi}) we introduced a functional
defined on functions {\bf f(k)} in the momentum representation:
\be
\J_{n,m}[{\bf f}]=
-i\frac{N_c}{(2\pi)^4}\int_{\Lambda^3} \frac{{\bf f(k)}d^4 k}{(m_u-k^2)^n(m_s-k^2)^m}.
\ee
This is an alternative to integrals $I_2^{f..f}$ which we thought 
better to introduce for a growing number of ``physical'' form factors.

We omit the intermediate calculation here.
For the decay constant $g_{K^{*'}\to K^*\pi}$  we find
	\be
	g_{K^{*'}\to K^*\pi}\approx 4\, {\rm GeV}^{-1}
	\ee
and the decay width is as follows:
	\be
	\Gamma_{K^{*'}\to K^*\pi}=
     {g_{K^{*'}\to K^*\pi}^2\over 32\pi M_{K^{*\prime}}^3}
	\Phi(M_{K^{*\prime}},M_{K^*},M_{\pi})^3\approx 90\mev.
	\ee
The lower limit for this value coming from
experiment is $\sim 91\pm 9$ MeV~\cite{PDG}.

A similar calculation has to be performed for the rest of the $K^{*'}$ decay
modes under consideration.
The coupling constant $g_{K^{*\prime}\to K\rho}$ is derived in the
same way as in (\ref{KstartoKstarpi}), with the only
difference that $\bar f_{\pi}$ and $\bar f_{K^*}$ are to be replaced
by $\bar f_{\rho}$ and $\bar f_{K}$.
The corresponding amplitude,
$T_{K^{*\prime}\to K\rho}^{\mu\nu}$, takes the form
	\be
	T_{K^{*'}\to K\rho}^{\mu\nu}= g_{K^{*'}\to K\rho}
	\epsilon^{\mu\nu\alpha\beta}p_{\alpha}q_{\beta},
	\label{T_Kstar_to_Krho}
	\ee
where $p$ and $q$ are   momenta of $K^{*'}$- and $K$-mesons,
respectively, and
	\be
	g_{K^{*'}\to K\rho}= {8m_s\over m_u^2-m_s^2}
	\left(
	\J_{2,0}[\bar f_{K^{*}}^\prime\bar f_K\bar f_{\rho}]-
	\J_{1,1}[\bar f_{K^{*}}^\prime\bar f_K\bar f_{\rho}]
	\right).  \label{g_Kstar_to_Krho}
	\label{KstartoKrho}
	\ee
The corresponding decay width is 
	\be
	\Gamma_{K^{*'}\to K\rho}
          = {g_{K^{*'}\to K\rho}^2\over 32\pi M_{K^{*\prime}}^3} 
	\Phi(M_{K^{*\prime}},M_K,M_\rho)^3.
	\ee
For the parameters given in Subsection~3.5 one has
\be
  g_{K^{*'}\to K\rho}\approx 3\, {\rm GeV}^{-1},\quad
\Gamma_{K^{*'}\to K\rho} \approx 20 \,{\rm MeV}.
\ee
From  experiment,  the upper limit for this process is
$\Gamma_{K^{*'}\to K\rho}^{exp} < 16\pm 1.5$ MeV.

The process $K^{*'}\to K\pi$ is described by the
amplitude
	\be
	T_{K^{*'}\to K\pi}^{\mu}=i \frac{g_{K^{*'}\to K\pi}}{2}
	(q-p)^{\mu},
	\ee
where $p$ and $q$ are  momenta of $\pi$ and $K$. 
The coupling constant $g_{K^{*'}\to K\pi}$ is obtained by calculating
the one-loop integral
	\be
	g_{K^{*'}\to K\pi}=
	4 \J_{1,1}[\bar f_{K^{*}}^\prime\bar f_{K}\bar f_{\pi}]
	\approx 2
	\label{KstartoKpi}
	\ee
and the decay width is
	\be
	\Gamma_{K^{*'}\to K\pi}
        ={g_{K^{*'}\to K\pi}^2\Phi(M_{K^{*\prime}},M_K,M_\pi)^3
\over 64\pi M^5_{K^{*'}}}\approx
	20\, {\rm MeV}.
	\ee
The experimental value is
$15\pm 5$ MeV~\cite{PDG}.

The mesons with hidden strangeness ($\varphi'$) are treated in
the same way as $K^{*'}$.
We consider  two decay modes: $\varphi'\to KK^*$ and
$\varphi'\to \bar KK$.
Their amplitudes are
	\ba
	T_{\varphi'\to KK^*}^{\mu\nu}&=& g_{\varphi'\to KK^*}
	\epsilon^{\mu\nu\alpha\beta}p_{\alpha}q_{\beta},\\
	T_{\varphi'\to \bar KK}^\mu&=& i g_{\varphi'\to \bar KK}
	(p-q)^{\mu}.
	\ea
Here,$p$ and $q$  are  momenta of the $K$- and $K^*$-mesons.
The related coupling constants are
	\ba
	g_{\varphi'\to KK^*}&=&
	{8m_u\over m_s^2-m_u^2}
	\left(
	 \J_{0,2}[\bar f_{\varphi}^\prime\bar f_{K^*}\bar f_{K}]-
	 \J_{1,1}[\bar f_{\varphi}^\prime\bar f_{K^*}\bar f_{K}]
	\right),\\
	g_{\varphi'\to \bar KK}&=&
	4 \J_{0,2}[\bar f_{\varphi}^\prime\bar f_{K}\bar f_{K}].
	\ea
Thus, the decay widths are estimated as
	\be
	\Gamma_{\varphi'\to KK^*}\approx 90 \, {\rm MeV}, \quad
	\Gamma_{\varphi'\to \bar KK}\approx 10 \, {\rm MeV}.
	\ee
Unfortunately,
there are no reliable experimental data  on the partial decay
widths for $\varphi'\to KK^*$ and $\varphi'\to \bar KK$
except the total width of $\varphi'$ being estimated
as $150\pm 50$ MeV~\cite{PDG}.
However, the dominance of the
process $\varphi'\to KK^*$ is observed
 is in agreement with our result.


Following the scheme outlined in the previous calculations,
we first estimate the $K'\to K^*\pi$ and $K'\to K\rho$ decay widths.
Their amplitudes are
	\ba
	T_{K'\to K^*\pi}^{\mu}&=&
	i g_{K'\to K^*\pi}(p+q)^{\mu},\\
	T_{K'\to K\rho}^{\mu}&=&
	i g_{K'\to K\rho}(p+q)^{\mu},
	\ea
here $p$ is the momentum of $K'$, $q$ is the momentum of $\pi$ ($K$).
 The coupling constants are
	\be
	g_{K'\to K^*\pi}=
	4\J_{1,1}[\bar f_{K}^\prime\bar f_{K^*}\bar f_{\pi}],\quad
	g_{K'\to K\rho}=
	4 \J_{1,1}[\bar f_{K}^\prime\bar f_{K}\bar f_{\rho}].
	\ee
By calculating the integrals in the  above
formulae we have $g_{K'\to K^*\pi}\approx-1.4$  and
$g_{K'\to K\rho}\approx-1.2$.
The decay widths thereby are
	\ba
	\Gamma_{K'\to K^*\pi}\approx 100\, \mev,\quad
	\Gamma_{K'\to K\rho}\approx 50\, \mev.
	\ea
These processes have been observed in experiment and
the decay widths are
\footnote{The accuracy
of measurements carried out for the decays of $K'$ is not given
in \cite{PDG}.}
 \cite{PDG}
	\be
	\Gamma^{exp}_{K'\to K^*\pi}\sim 109\, \mev,\quad
	\Gamma^{exp}_{K'\to K\rho}\sim 34\, \mev.
	\ee

The remaining  decay $K'\to K\pi\pi$  into three particles
requires more complicated calculations.
In this case, one must consider a box diagram, Fig.\ref{KtoK2pi}.(a),
and two types of diagrams, Fig.\ref{KtoK2pi}.(b), with
intermediate $\sigma-$ and $K^*_0-$resonances.
The diagrams for resonance channels are approximated by
 the relativistic Breit-Wigner function.
The integration over the kinematically relevant range in
the phase space for final states gives
	\be
	\Gamma_{K'\to K\pi\pi}\sim 1 {\rm MeV}.
	\ee

\subsection{Strong decays of  scalar mesons}
The ground and excited states of scalar mesons $f_0$, $a_0$, and $\kstar$
decay mostly into  pairs of pseudoscalar
mesons.  

They can easily be related to $Z^a_{\sigma(\varphi),i}$ introduced at the
beginning of our paper.

All amplitudes that describe processes of the type $\sigma\to\varphi_1\varphi_2$
can be divided into two parts:
\ba
	T_{\sigma\to\varphi_1\varphi_2}&=&
	C\left(-\frac{i N_c}{(2\pi)^4}\right)
	\int_{\Lambda_3}d^4 k \frac{\Tr[(m+\Slash{k}+\Slash{p}_1)\gamma_5
	(m+\Slash{k})\gamma_5(m+\Slash{k}-\Slash{p}_2)]}{
	(m^2-k^2)(m^2-(k+p_1)^2)(m^2-(k-p_2)^2)}\nonumber\\
	&=& 4mC\left(-\frac{i N_c}{(2\pi)^4}\right)
	\int_{\Lambda_3}d^4k
	\frac{\left[1-\displaystyle\frac{p_1\cdot p_2}{m^2-k^2}\right]}{(m^2-(k+p_1)^2)(m^2-(k-p_2)^2)}
	\nonumber\\
	&=&4 m C [I_2(m,p_1,p_2)-p_1\cdot p_2 I_3(m,p_1,p_2)]=T^{(1)}+T^{(2)}.
	\label{T}
\ea
Here $C=4 g_{\sigma} g_{\varphi_1}g_{\varphi_2}$ and $p_1$, $p_2$ are 
momenta of  pseudoscalar mesons.
We rewrite the amplitude
$T_{\sigma\to\varphi_1\varphi_2}$
in another form
\ba
&&T_{\sigma\to\varphi_1\varphi_2}\approx
4mZ^{-1/2}g_{\varphi_1} \left[1-p_1\cdot p_2
\frac{I_3(m)}{I_2(m)}\right],\label{T1}\\
&&p_1\cdot p_2 =\frac12(M_\sigma^2-M_{\varphi_1}^2-M_{\varphi_2}^2).
\ea
We assumed here that the  $I_3/I_2$ ratio  slowly changes with the momentum
in comparison with the factor $p_1\cdot p_2$,
therefore, we ignore their momentum dependence in (\ref{T1}).
With this assumption we
are going to obtain just a qualitative picture for  decays of the
excited scalar mesons.

In Eqs. (\ref{T}) and (\ref{T1}), we omitted the contributions
from the diagrams that include form factors at vertices.
The whole set of diagrams consists of those containing zero, one, two,
and three form factors. To obtain the complete amplitude, one must
sum up all contributions.

After these general comments, let us  consider the decays of
$ a_0(1450)$, $f_0(1370)$, $f_J(1710)$, and $\kstar(1430)$.
First, we estimate the
decay width of the process $\hat a_0\to\eta\pi$,
taking the mixing coefficients from Tables~\ref{mixingTable} and~\ref{isoscMixTab}
(see Appendix B for details).
The result is
\be
      T^{(1)}_{\hat a_0\to\eta\pi}\approx0.2\gev,\quad
     T^{(2)}_{\hat a_0\to\eta\pi}\approx3.5 \gev,
\ee
\be
     \Gamma_{\hat a_0\to\eta\pi}\approx
     160 \mev.
\ee

From this calculation one can see that $T^{(1)}\ll T^{(2)}$ and the
amplitude is dominated by its second part, $T^{(2)}$, that is
momentum-dependent. The first part is small because the diagrams
with different numbers of form factors cancel each other. As a consequence,
in all processes where an excited scalar meson decays into a pair of
ground pseudoscalar states, the second part of the amplitude
determines the rate of the process.

For the decay $\hat a_0\to\pi\eta'$ we obtain the amplitudes
\be
     T^{(1)}_{\hat a_0\to\pi\eta'}\approx0.8 \gev,\quad
     T^{(2)}_{\hat a_0\to\pi\eta'}\approx3 \gev,
\ee
and the decay width
\be
    \Gamma_{\hat a_0\to\pi\eta'}\approx36 \mev.
\ee
The decay of $\hat a_0$ into kaons is described by the amplitudes $T_{\hat a_0\to K^+K^-}$
and  $T_{\hat a_0\to \bar K^0K^0}$
which, in accordance with our scheme,  can again be  divided into two parts: $T^{(1)}$
and $T^{(2)}$ (see Appendix B for details):
\be
T_{\hat a_0\to K^+K^-}^{(1)}\approx 0.2\gev,\quad
T_{\hat a_0\to K^+K^-}^{(2)}\approx 2.1\gev
\ee
and the  decay width  is
\be
\Gamma_{\hat a_0\to KK}=\Gamma_{\hat a_0\to K^+K^-}+\Gamma_{\hat a_0\to \bar K^0K^0}\approx 100\mev.
\ee
Qualitatively, our results do not contradict the experimental data.
\be
	\Gamma^{\rm tot}_{\hat a_0}=265\pm13 \mev,\quad BR(\hat a_0\to KK):BR(\hat a_0\to\pi\eta)= 0.88\pm0.23.
\ee
The decay widths
of radial excitations of scalar isoscalar mesons
are estimated in the same way as shown above:
\be
\Gamma_{\hat\sigma\to\pi\pi}\approx\left\{
\begin{array}{l}
550 \mev (M_{\hat\sigma}=1.3 \gev) \\
460 \mev (M_{\hat\sigma}=1.25 \gev),
\end{array}
\right.
\ee
\be
\Gamma_{\hat\sigma\to\eta\eta}\approx\left\{
\begin{array}{l}
24 \mev (M_{\sigma}=1.3 \gev) \\
15 \mev (M_{\sigma}=1.25 \gev),
\end{array}
\right.
\ee
\be
\Gamma_{\hat\sigma\to\sigma\sigma}\approx\left\{
\begin{array}{l}
6 \mev (M_{\sigma}=1.3 \gev) \\
5 \mev (M_{\sigma}=1.25 \gev),
\end{array}
\right.
\ee
\be
\Gamma_{\hat\sigma\to KK}\sim 5 \mev,
\ee
\be
\Gamma_{\kstar\to K\pi}\approx 300\mev.
\ee


The heaviest scalar isoscalar meson in our model has the mass $1600\mev$ 
(see Table~\ref{masses}) to be associated with an experimentally
found meson state.  From experimental data \cite{PDG}, we find two
possible candidates for the role of a member of the radially 
excited meson nonet: $f_0(1500)$ and $f_0(1710)$. 
The extra meson state can be explained by possible mixing of 
 members of the $\bar qq$ meson nonets with
a  gluon bound state, the glueball.
Indeed, on the mass scale, 
both  meson states lie in the region where the hypothetical glueball state
is expected to exist.  Insofar as we did not include  the glueball 
into our model (however, we are going to do this in our further works), the
picture is not complete. Nevertheless, we are free to  make a hypothesis
concerning the contents of $f_0(1500)$ and $f_J(1710)$.  We expect
that one of these states is mostly a quarkonium with just a
negligible admixture of the glueball state whereas the other
is essentially mixed with the glueball. The mass splitting that always
appears when two or more states mix with each other will ether increase or decrease
the mass of a quarkonium, depending on the mass of a ``bare'' (unmixed)
glueball state either being smaller or greater than the mass of the quarkonium. 
After mixing we expect to find the $\bar qq$ bound state with the
mass $1500\mev$ or $1710\mev$.  

To decide which of them is the 
quarkonium with a small content of a glueball state, associated with
the radial excitation of $f_0(980)$, 
we estimate its decay widths for two cases: first for  the
mass $1710\mev$ quarkonium
\be
\begin{array}{lclclcl}
\Gamma_{f_0(1710)\to 2\pi}& \approx& 3\mev, &\quad &
\Gamma_{f_0(1710)\to 2\eta}& \approx& 40\mev, \\
\Gamma_{f_0(1710)\to \eta\eta'}& \approx& 42\mev, &\quad &
\Gamma_{f_0(1710)\to KK}& \approx& 24\mev, 
\end{array}
\ee
and then for the mass $1500\mev$ quarkonium
\be
\begin{array}{lclclcl}
     \Gamma_{f_0(1500)\to 2\pi}&\approx& 3\mev, &\quad &
     \Gamma_{f_0(1500)\to 2\eta}&\approx& 20\mev,\\
     \Gamma_{f_0(1500)\to \eta\eta'}&\approx& 10\mev, &\quad &
     \Gamma_{f_0(1500)\to KK}&\approx& 20\mev.
\end{array}
\ee
The decays of $f_0(1500)$ and $f_0(1710)$ into $\sigma\sigma$ are negligible,
so we disregard them.
From the experimental data we have:
\be
\Gamma^{\rm tot}_{\sigma'}=200 - 500 \mev,
\quad
\Gamma^{\rm tot}_{f_0(1710)}=133\pm 14 \mev,
\quad
\Gamma^{\rm tot}_{f_0(1500)}=112\pm 10 \mev.
\ee
Thus, we can see that in the case of $f_0(1500)$ being a $\bar qq$ state
there is a deficit in the decay widths whereas for $f_J(1710)$ the result is
close to experiment.
From this we conclude that  the  meson
$f_J(1710)$ better suits for the role of
a member of the  $\bar qq$ nonets as
a radially excited partner for $f_0(980)$
and the meson state $f_0(1370)$ as the
first radial excitation  of $f_0(400-1200)$.
As to  $f_0(1500)$,  the $\bar qq$ model works bad 
for it. This gives us the idea that $f_0(1500)$ 
is essentially mixed with the glueball state which
significantly contributes to its decay width.
 Our interpretation of $f_0(1500)$ and $f_0(1710)$ is in
agreement with other approaches where similar conclusions were
made by the $K$-matrix method \cite{anisovich} and
QCD sum rules \cite{narison}.


The strong decay widths of  ground states of scalar mesons
were calculated in paper \cite{volk98} in the framework of 
the standard NJL model with 't Hooft interaction where it was
shown that a strange scalar meson state with a mass about $960\mev$ 
decays into $K\pi$ with the rate
\ba
&&\Gamma_{\kstar(960)\to K\pi}=
\frac{3}{Z\pi M_{\kstar}^3}\left(\frac{m_u m_s}{2F_\pi}\right)^2
\Phi(M_{\kstar},M_{K},M_\pi)
\approx360 \mev.
\ea
By comparing this result with the analysis of phase shifts given in
\cite{ishida} where  an evidence for existence of a scalar strange
meson with the mass equal to $905\pm50\mev$ and decay width 
$545\pm170\mev$ is shown, we identify the state $\kstar(960)$ as
a member of the ground scalar meson nonet. The state $\kstar(1430)$
is thereby its first radial excitation.

\subsection{Strong decays of $\eta(1295)$ and $\eta(1440)$.}

The mesons $\eta(1295)$ and $\eta(1440)$ have common
decay modes: $a_0\pi$, $\eta\pi\pi$, $\eta(\pi\pi)_{\sl S-wave}$,
$K\bar K\pi$,
moreover, the heavier pseudoscalar $\eta(1440)$ decays also into
$KK^*$. For the processes with two secondary particles,
the calculations of decay widths are done in the same way as
shown in the previous subsection, by calculating 
the corresponding triangle diagrams.

Let us consider the decay $\eta\to a_0\pi$. The corresponding
amplitude is of the same form  as given in (\ref{T}) for
decays of the type $\sigma\to\varphi\varphi$.
It can also be divided into two parts $T^{(1)}$ and  $T^{(2)}$
which in our approximation
are constant and momentum-dependent in the sense explained in
the previous subsection (see (\ref{T1}) and the text below):
\be
T^{(1)}_{\hat\eta\to a_0\pi}\approx 0.3 \gev,\quad
T^{(2)}_{\hat\eta\to a_0\pi}\approx -1  \gev
\ee
Therefore, the decay width is
\be
\Gamma_{\hat\eta\to a_0\pi}\approx 3 \mev.
\ee

The decay $\hat\eta\to \eta(\pi\pi)_{\sl S-wave}$ is nothing else than
the decay $\hat\eta\to\eta\sigma\to\eta(\pi\pi)_{\sl S-wave}$ where
we have the $\sigma$-meson in the final state decaying then into
pions in the S-wave. We simply calculate $\hat\eta\to\eta\sigma$,
with $\sigma$ as a  decay product.

\begin{table}
\caption{$\eta(1295)$ and $\eta(1440)$ decay modes. }
\label{etadecay}
\begin{tabular}{||c|c|c|c|c|c|c||}
\hline\hline
 &   $a_0\pi$ & $\eta\sigma$ & $\eta\pi\pi$ & $K\bar K\pi$ & $KK^*$&
$\Gamma^{\rm tot}$\\
\hline
$\eta(1295)$ &  $3\mev$  & $30\mev$  & $4\mev$& $5\mev$ & $-$ & 48\mev \\
\hline
$\eta(1440)$ &  $10\mev$  & $3\mev$  & $6\mev$& $26\mev$ & $70~{\rm keV}$ &45\mev\\
\hline
\end{tabular}
\end{table}

The calculation of decay widths for the rest of the decay modes with
two particles in the final state is similar and the result
is given in Table~\ref{etadecay}.

The decay $\hat\eta'\to KK^*$ differs from the other modes
due to the strange vector meson among the decay products.
In this case we have
\ba
     T_{\hat\eta'\to KK^*}^\mu&=&
     4(p_1+p_2)^\mu \biggl([g_{u}g_{K}g_{K^*}I_2(m_u,m_s)+\ldots]-\nonumber\\
     &&\sqrt{2}[g_{s}g_{K}g_{K^*}I_2(m_u,m_s)+\ldots]\biggr)
\ea
where $p_1$ is the momentum of $\hat\eta'$; $p_2$, the momentum of
$K$; and dots stand for the terms with form factors (not displayed
here). 
These two parts are of the same order of magnitude and differ in 
sign and therefore cancel each other, which 
reduces the decay width up to tens of keV:
\be
\Gamma_{\hat\eta'\to KK^*}\approx 70~{\rm keV}.
\ee

When there are three particles in the final state, poles appear
in amplitudes, related to intermediate scalar
resonances. As it is well known from $\pi\pi$  scattering,
these diagrams can play a crucial role in the description of
such processes.
So, in addition to the "box" diagram we take  account  of the  diagrams with
poles provided by $\sigma$, $f_0$, and $a_0$ resonances
(see Fig.~\ref{polediag}).
Here we neglect the momentum dependence 
in the box diagram approximating it by a constant.
The amplitude is thereby
\ba
&&T_{\hat\eta\to\eta\pi\pi}=
B+{c_{\sigma\eta\hat\eta} c_{\sigma\pi\pi} \over
     M_{\sigma}^2-s-i M_{\sigma}\Gamma_{\sigma}   }+
{c_{f_0\eta\hat\eta} c_{f_0\pi\pi} \over
     M_{f_0}^2-s-i M_{f_0}\Gamma_{f_0}   }\nonumber\\
&&\quad+{c_{a_0\hat\eta\pi} c_{a_0\eta\pi} \over
     M_{a_0}^2-t-i M_{a_0}\Gamma_{a_0}   }+
{c_{a_0\hat\eta\pi} c_{a_0\eta\pi} \over
     M_{a_0}^2-u-i M_{a_0}\Gamma_{a_0}   }+  excited,
\ea
where $B$ is given by the  "box" diagram:
\be
B= 12 \left(\frac{m_u}{F_\pi}\right)^2Z^{-1}[R_{11}R_{12}+\ldots]
\ee
where dots stand for the contribution from diagrams with form factors, 
and $R_{ij}$ are taken from Table~\ref{isoscMixTab} (for $\eta$ and $\hat\eta$).
The coefficients $c_{\sigma\varphi\varphi}$ represent
the amplitudes describing decays of a scalar to a couple of
pseudoscalars; the calculation of them 
was discussed in the previous subsection. 
In general, they are momentum-dependent.

The kinematic invariants $s$, $t$, and $u$ are Mandelstam variables:
$s=(p_{\pi_1}+p_{\pi_2})^2$, $t=(p_{\eta}+p_{\pi_1})^2$,
$u=(p_{\eta}+p_{\pi_2})^2$

The ``{\it excited\/}''
terms are contributions from excited scalar resonances
of a structure similar to that for the ground states.
The decay widths of  processes $\hat\eta\to \eta\pi\pi$ and
$\hat\eta'\to\eta\pi\pi$ are thereby
\be
\Gamma_{\hat\eta\to\eta\pi\pi}\approx 4\mev,\quad
\Gamma_{\hat\eta'\to\eta\pi\pi}\approx 6\mev.
\ee

For the processes $\hat\eta\to K\bar K\pi$
and $\hat\eta' \to K\bar K\pi$ we
approximate their decay widths by neglecting the pole-diagram
contribution
because it turns out that the "box" is dominant here.
The result is given in Table~\ref{etadecay}.

Unfortunately, the branching ratios for different decay modes 
of $\eta(1295)$ and $\eta(1440)$ are not  well known from
experiment; so one can only find their total decay widths
\be
\Gamma^{\rm tot}_{\eta(1295)}=53\pm6 \mev,\quad
\Gamma^{\rm tot}_{\eta(1440)}=50-80 \mev,
\ee
which is in satisfactory agreement  with our results.

Strong and electromagnetic decays of the ground states of $\hat\eta$
and $\hat\eta'$ mesons 
were  investigated within the framework of 
the standard NJL model in \cite{volk_84, volk_86} and we do not consider them
here.

\section{Conclusion}

Let us summarize and discuss 
main features of the  nonlocal 
NJL model proposed here and basic results  obtained in our work.

A simple generalization of the NJL model to a nonlocal 
four-quark interaction of the separable type was suggested to
describe first radial excitations of the scalar, pseudoscalar,
and vector mesons. The nonlocality was introduced
into quark currents by means of simple form factors,
while preserving the local form of the ground and excited
meson states. On the one hand, form factors can be written in a relativistic
invariant form. On the other hand, the form
factor parameters can be chosen so that the gap equations 
keep the conventional form, which leads to constant 
constituent quark masses and quark condensates.
As a result, all low energy theorems are fulfilled in 
our model in the chiral limit (see Section~2).
Therefore, the introduction of excited meson states does not
destroy those attractive features which the NJL model is
characteristic of.

The model contains six basic and seven additional 
form factor parameters. The basic ones are defined like in
the standard (local) NJL model. They are the quark masses
$m_u=m_d$, $m_s$, the cut-off parameter $\Lambda_3$, 
and three quark coupling constants $G$, $G_V$, $K$.
To determine them, we used six input quantities:
$F_\pi$, $g_\rho$, $M_\pi$, $M_K$, $M_\rho$, and the 
mass difference $M_{\eta}^2-M_{\eta'}^2$. Then,
we predicted the masses of $\eta$, $\eta'$, $\kstar$,
$\varphi$ mesons and also the masses of the scalar and
axial-vector meson nonets. The weak decay constant $F_K$
and all strong coupling meson constants are calculated.

Upon the  excited meson states are included,
a great number of form factors appears in the model.
They are necessary to describe radial excitations of
the three meson nonets: scalar, pseudoscalar and
vector. Each form factor contains two parameters:
the external parameter $c_{qq}^U$ characterizing
to what extent the interaction of excited states is 
stronger than that of the ground ones and 
the internal (slope) parameter $d_{qq}$
determining the shape of the wave function of an excited meson state.

We give an unambiguous definition of
the slope parameters for  scalar mesons from the condition 
that the excited states do not contribute to quark condensates.
Then, we assume the slope parameters to be the same for 
any sort of meson fields. Moreover, in favor of the global
chiral symmetry,  we  put the scalar meson form factors 
equal to the pseudoscalar meson ones.
As a result, only seven independent parameters are left:
$c^{\pi}_{uu}$, $c^K_{us}$, $c^{\eta,\eta'}_{uu}$, $c^{\eta,\eta'}_{ss}$,
$c^{\rho,\omega}_{uu}$, $c^{\kstar}_{us}$, $c^{\varphi}_{ss}$.
They are fixed by masses of radially excited pseudoscalar and
vector mesons. When this procedure is completed, we are able
to  predict the masses of  scalar mesons  and identify 
them with experimentally observed meson states.

The major results obtained in our work are:
\begin{itemize}
\item[1)] A nonlocal chiral quark model with a quark interaction of the 
separable type   was developed to describe the ground and first radially excited states 
of mesons represented by local fields. In this model, 
the quark condensate and  gap equations are conserved in the standard form, and 
all low-energy theorems are fulfilled.
\item[2)] In a realistic $U(3)\times U(3)$ version of the model, the 
$U_A(1)$ problem is solved by introducing the 't Hooft interaction.
The mixing of  pseudoscalar isoscalar meson states, the 
ground $\eta$, $\eta'$, and the radially excited $\hat\eta$, $\hat\eta'$,
due to the 't Hooft interaction, was taken into account.
\item[3)] In the framework of the proposed model, a satisfactory description of
the masses of  ground and first radially excited 
pseudoscalar and vector meson states was obtained.
\item[4)] The mass spectrum for  scalar meson nonets (ground and
first radially excited) is predicted on the basis of the 
proposed model and with the assumption on the form factors, based on 
the global chiral symmetry, that the form factors for scalar mesons are
the same as for the pseudoscalars ones.    
\item[5)] The members of  quark-antiquark
nonets, whose physics the proposed model is intended to describe,  
are identified with 
twenty seven physically observed scalar, pseudoscalar, and vector meson states,
\item[6)] The weak decay constants $F_{\pi'}$, $F_K$, and $F_{K'}$
are estimated.
\item[7)] The widths of  main strong decays of radially excited
scalar, pseudoscalar, and vector meson nonets 
are estimated. The results are in satisfactory agreement with experimental data.
\end{itemize}

Let us make some comments on the identification of 
the meson nonets' members. 
While it seems clear how to identify the members of 
pseudoscalar and vector meson nonets, the scalar mesons require more words to say.
From our calculations we come to the following interpretation of
$f_0(1370)$, $f_J(1710)$, $a_0(1470)$, $\kstar(1430)$ mesons: we consider them
as the first radial excitations of the ground states
$f_0(400-1200)$, $f_0(980)$, $a_0(980)$ and $\kstar(960)$
\footnote{
The light strange scalar of a mass about 900 \mev~is 
not included into the summary tables of PDG \cite{PDG}. However, 
there are evidences from the phase shift analysis \cite{ishida} that 
a state (known as $\kappa(900)$) with the mass 950\mev~does exist.
}. 

In this picture, however, no place is reserved for the $f_0(1500)$
meson. To include it,  we need an additional
meson state in our model that is not a bound $\bar qq$-system
(there is no vacancy in the considered multiplets) but rather it is
a bound colorless gluon state \cite{glueball}. 
There are many reasons that the state $f_0(1500)$ is essentially
mixed with a glueball \cite{anisovich,narison}. However, in
this paper we did not take the glueball into account. Therefore,
we cannot say how much it can affect $\bar qq$ meson states. 
However, we are going to tackle this problem in our
further work. In the present paper, we obtain a bound quark-antiquark
state with the mass about 1600\mev, so we have to decide 
which of the observed meson states, $f_0(1500)$ or $f_J(1710)$,
is to be associated with this member of the nonet of the radially excited scalar 
mesons in our model. We have chosen $f_J(1710)$. The reason for
this choice is based both 
upon the results obtained in ref.~\cite{anisovich,narison}
and on our estimates of the decay widths  discussed in Section 4.

In conclusion, we would like to outline further steps
to improve  our model. First of all, a glueball state
can be included into the effective Lagrangian.
This will allow us to correct the description of the scalar states $f_0(980)$,
$f_0(1370)$, $f_0(1710)$ and include $f_0(1500)$ (presumed to be
essentially mixed with a glueball) into the whole picture. The mixing of
all the states will play an important role in this case. By now,
we took  account  only of the mixing among  $f_0(400-1200)$, $f_0(980)$,
$f_0(1370)$, $f_0(1710)$ and among $\eta$, $\eta'$, $\hat\eta$,
$\hat\eta'$. Nevertheless, our investigation revealed that the
meson states $\eta(1300)$, $\eta(1470)$, $f_0(1370)$, $a_0(1470)$,
$f_0(1710)$, $\kstar(1430)$ are the first radial excitations of
 $\eta(590)$, $\eta(950)$, $f_0(400-1200)$, $a_0(980)$,
$f_0(980)$, $\kstar(960)$.

Second, the absence of quark confinement is still a common flaw of 
NJL-like models with a local quark interaction. There are several
approaches suggested to find a solution of this problem. Among them there are
various potential models, models where the pole in the quark propagator is
excluded \cite{efr96}, {\it etc.\/} We are going to continue to work with our
own approach which was suggested in ref. \cite{conf}. 
\vspace{1cm}

\centerline{\Large\bf Acknowledgment}

We would like to thank  Dr.~C.~Weiss who made a large contribution to
fundamentals of this model and also our collaborators Prof.~D.Ebert and
Dr.~M.~Nagy. We are grateful also to Prof.~S.B.~Gerasimov for fruitful
discussions. This work has been supported by RFBR Grant 98-02-16185.



\newpage
\appendix

\section*{Appendix}

\section{Coefficients of the free part of the effective Lagrangian
for  scalar isoscalar mesons.}

The functions $K_{\sigma(\varphi),ij}^{[a,b]}$ introduced in Sec.~3 of
Chap.~4 (\ref{Lisosc})
are defined as follows
\ba
     K_{\sigma(\varphi),11}^{[a,a]}(P)&=&
     Z_{\sigma(\varphi),1}^a (P^2 -
     (m_q^a\pm m_{q'}^a)^2-  M_{\sigma^a(\varphi^a),1}^2 ),\nonumber\\
     K_{\sigma(\varphi),22}^{[a,a]}(P)
	 &=& Z_{\sigma(\varphi), 2}^a (P^2 -
     (m_q^a\pm m_{q'}^a)^2-  M_{\sigma^a(\varphi^a),2}^2 ),
     \nonumber \\
     K_{\sigma(\varphi),12}^{[a,a]}(P) &=& K_{\sigma(\varphi),21}^{[a,a]}(P) \;\; = \;\;
     \gamma_{\sigma(\varphi)}^a (P^2 - (m_q^a\pm m_{q'}^a)^2 ),\\
     K_{\sigma(\varphi),11}^{[8,9]}(P)&=&K_{\sigma(\varphi),11}^{[9,8]}(P)=
	\frac12 \left(T^{S(P)}\right)^{-1}_{89},\nonumber\\
	K_{\sigma(\varphi),12}^{[8,9]}(P)&=&
	K_{\sigma(\varphi),12}^{[9,8]}(P)=
	K_{\sigma(\varphi),21}^{[8,9]}(P)=0,\nonumber\\
	K_{\sigma(\varphi),21}^{[9,8]}(P)&=&
	K_{\sigma(\varphi),22}^{[8,9]}(P)=
	K_{\sigma(\varphi),22}^{[9,8]}(P)=0 \nonumber
\ea
where the ``bare'' meson masses are
\ba
     && M_{\sigma^8(\varphi^8),1}^2= (Z^8_{\sigma(\varphi),1})^{-1}
	\left({1\over 2}(T^{S(P)})^{-1}_{88} - 8I_1(m_u) \right),
     \nonumber \\
     && M_{\sigma^9(\varphi^9),1}^2= (Z^9_{\sigma(\varphi),1})^{-1}
	\left({1\over 2}(T^{S(P)})^{-1}_{99}-8I_1(m_s) \right), \nonumber\\
     && M_{\sigma^8(\varphi^8),2}^2=(Z^8_{\sigma(\varphi),2})^{-1}
	\left({1\over 2G} - 8I_1^{ff}(m_u) \right),\\
     && M_{\sigma^9(\varphi^9),2}^2=(Z^9_{\sigma(\varphi),2})^{-1}
	\left({1\over 2G}-8I_1^{ff}(m_s) \right). \nonumber
     \label{Mpuu}
\ea
In the case of isoscalar mesons it is convenient to combine the scalar and pseudoscalar
fields into 4-vectors
\be
     \Phi=(\varphi_{1}^{8\,r},\varphi_{2}^{8\,r},
	\varphi_{1}^{9\,r},\varphi_{2}^{9\,r}), \qquad
     \Sigma=(\sigma_{1}^{8\,r},\sigma_{2}^{8\,r},
	\sigma_{1}^{9\,r},\sigma_{2}^{9\,r}),
\ee
and introduce  $4\times 4$ matrix functions ${\cal K}_{\sigma(\varphi),ij}$,
instead of old $K_{\sigma(\varphi),ij}^{[a,b]}$,
where indices $i,j$ run from 1 through 4. This allows us to
rewrite the free part of the effective Lagrangian which then, with
the meson fields renormalized, looks
as follows
\ba
     {\cal L}^{(2)}(\Sigma,\Phi)=\frac12\sum_{i,j=1}^4
     \left(\Sigma_i {\cal K}_{\sigma,ij}(P)\Sigma_j+
     \Phi_i {\cal K}_{\varphi,ij}(P)\Phi_j\right).
	\label{newL2}
\ea
and the functions  ${\cal K}_{\sigma(\varphi),ij}$ are
\ba
     {\cal K}_{\sigma(\varphi),11}(P)&=&P^2-(m_u\pm m_u)^2-M_{\sigma^8(\varphi^8),1}^2,\nonumber\\
     {\cal K}_{\sigma(\varphi),22}(P)&=&P^2-(m_u\pm m_u)^2-M_{\sigma^8(\varphi^8),2}^2,\nonumber\\
     {\cal K}_{\sigma(\varphi),33}(P)&=&P^2-(m_s\pm m_s)^2-M_{\sigma^9(\varphi^9),1}^2,\nonumber\\
     {\cal K}_{\sigma(\varphi),44}(P)&=&P^2-(m_s\pm m_s)^2-M_{\sigma^9(\varphi^9),2}^2,\\
     {\cal K}_{\sigma(\varphi),12}(P)&=&{\cal K}_{\sigma(\varphi),21}(P)=\Gamma_{\sigma_u(\eta_u)}(P^2-(m_u\pm m_u)^2),\nonumber\\
     {\cal K}_{\sigma(\varphi),34}(P)&=&{\cal K}_{\sigma(\varphi),43}(P)=\Gamma_{\sigma_s(\eta_s)}(P^2-(m_s\pm m_s)^2),\nonumber\\
     {\cal K}_{\sigma(\varphi),13}(P)&=&{\cal K}_{\sigma(\varphi),31}(P)=
	(Z^8_{\sigma(\varphi),1}Z^9_{\sigma(\varphi),2})^{-1/2}(T^{S(P)})^{-1}_{89}. \nonumber
\ea
Now,  to transform (\ref{newL2}) to the conventional form, one
should just diagonalize a 4-dimensional matrix, which is better to
do numerically.

\section{The calculation of the amplitudes for  decays of the
excited scalar meson $\hat a_0$}
Here we collect some instructive formulae that display a part of the
details of calculations made in this work. Let us demonstrate
how the amplitude of the decay $\hat a_0\to\eta\pi$ is obtained.
The mixing coefficients are taken from  Table~\ref{mixingTable}.
Moreover, the diagrams where  pion vertices contain form factors are
neglected because, as one can see from Table~\ref{mixingTable},
their contribution is significantly reduced:
\ba
      T^{(1)}_{\hat a_0\to\eta\pi}&=&4 \frac{m_u^2}{F_\pi}\biggl\{
	0.82\cdot0.71\cdot Z^{-1/2}\frac{I_2(m_u)}{I_2(m_u)}-\nonumber\\
	&&\left(1.17\cdot 0.71\cdot Z^{-1/2}-0.82\cdot 0.11\right)
	\frac{I_2^f(m_u)}{\sqrt{I_2(m_u)I_2^{ff}(m_u)}}-\nonumber\\
	&&1.17\cdot0.11\cdot\frac{I_2^{ff}(m_u)}{I_2^{ff}(m_u)}\biggr\}\approx0.2\gev,
\ea
\ba
     T^{(2)}_{\hat a_0\to\eta\pi}&=&
     2\frac{m_u^2}{F_\pi}(M_{a_0}^2-M_{\eta}^2-M_{\pi}^2)
     \biggl\{
        0.82\cdot0.71 Z^{-1/2}\frac{I_3(m_u)}{I_2(m_u)}-\nonumber\\
	&&\left(1.17\cdot 0.71\cdot Z^{-1/2}-0.82\cdot 0.11\right)
	\frac{I_3^f(m_u)}{\sqrt{I_2(m_u)I_2^{ff}(m)}}-\nonumber\\
	&&1.17\cdot0.11\frac{I_3^{ff}(m_u)}{I_2(m_u)}\biggr\}\approx3.5 \gev.
\ea
The decay width  thereby is
\be
     \Gamma_{\hat a_0\to\eta\pi}=
     \frac{|T_{\hat a_0\to\eta\pi}|^2}{16\pi M_{\hat a_0}^3}
     \sqrt{M_{\hat a_0}^4\!+\!M_{\eta}^4\!+\!M_{\pi}^4\!-\!
     2(M_{\hat a_0}^2 M_{\eta}^2\!+\!M_{\hat a_0}^2 M_{\pi}^2\!+\!M_{\eta}^2 M_{\pi}^2)}\approx
     160 \mev.
\ee
Here $I_2(m_u)=0.04$, $I_2^f(m_u)=0.014c$, $I_2^{ff}(m_u)=0.015c^2$,
$I_3(m_u)=0.11 \gev^{-2}$, $I_3^f(m_u)=0.07c\gev^{-2}$,$I_3^{ff}(m_u)=0.06c^2\gev^{-2}$
and $c$ is the external form factor parameter factored out
and cancelled in the ratios of  integrals.

For the decay into strange mesons we obtain (see Fig.1)
\ba
     &&\!\!\!\!T_{\hat a_0\to K^+K^-}\!=\!
     C_K\!\left(-\frac{iN_c}{16\pi^2}\right)\!\!\int\!d^4k
	{ \Tr[(m_u+\Slash{k}+\Slash{p}_1)\gamma_5(m_s+\Slash{k})\gamma_5(m_u+\Slash{k}-\Slash{p}_2)]
      \over
	(m_s^2-k^2)(m_u^2-(\Slash{k}-\Slash{p}_1)^2)(m_u^2-(\Slash{k}-\Slash{p}_2)^2)
	}\approx\nonumber\\
     &&\!\!\!\!2C_K\left\{
	 (m_s+m_u)I_2(m_u)-\Delta I_2(m_u,m_s)-
	 [m_s(M_{\hat a_0}^2-2M_{K}^2)-\right.\\
     &&\!\!\!\!\left.\quad	2\Delta^3]I_3(m_u,m_s)\nonumber
        \right\},
\ea
where $\Delta=m_s-m_u$ and
\be
	I_3(m_u,m_s)=-i\frac{N_c}{(2\pi)^4}\int_{\Lambda_3}\!\frac{d^4k}{(m_u^2-k^2)^2(m_s^2-k^2)}.
\ee
The coefficient $C_K$ absorbs the Yukawa coupling constants and some structure coefficients.
The integral $I_2(m_u,m_s)$ is defined by (\ref{DefIf}).
This is only the part of the amplitude without form factors.
The complete amplitude of this process is a sum of contributions which
contain also the integrals $I_2^{f..f}$ and $I_3^{f..f}$ with form factors.
Thus, the amplitude is
\ba
	&&T_{\hat a_0\to K^+K^-}=T^{(1)}+T^{(2)},\\
	&&T^{(1)}=\frac{m_u+m_s}{2F_K}
	\bigl\{
	(m_s+m_u)\cdot 0.13-\Delta\cdot 0.21\bigr\}\approx 0.2\gev,\\
	&&
	T^{(2)}=\frac{m_u+m_s}{2F_K}\bigl\{[m_s(M_{a_0}^2-2M_{K}^2)-2\Delta^3]\cdot1\gev^{-2}\bigr\}\approx 2.3\gev
	,\\
	&& F_K=1.2F_\pi.\nonumber
\ea
The decay width therefore is evaluated to be
\be
	\Gamma_{\hat a_0\to K^+K^-}=\Gamma_{\hat a_0\to \bar K^0K^0}\approx 50 \mev.
\ee




\clearpage

\clearpage
\section*{Figure captions}
\begin{figure}[t]
\centerline{\psfig{file=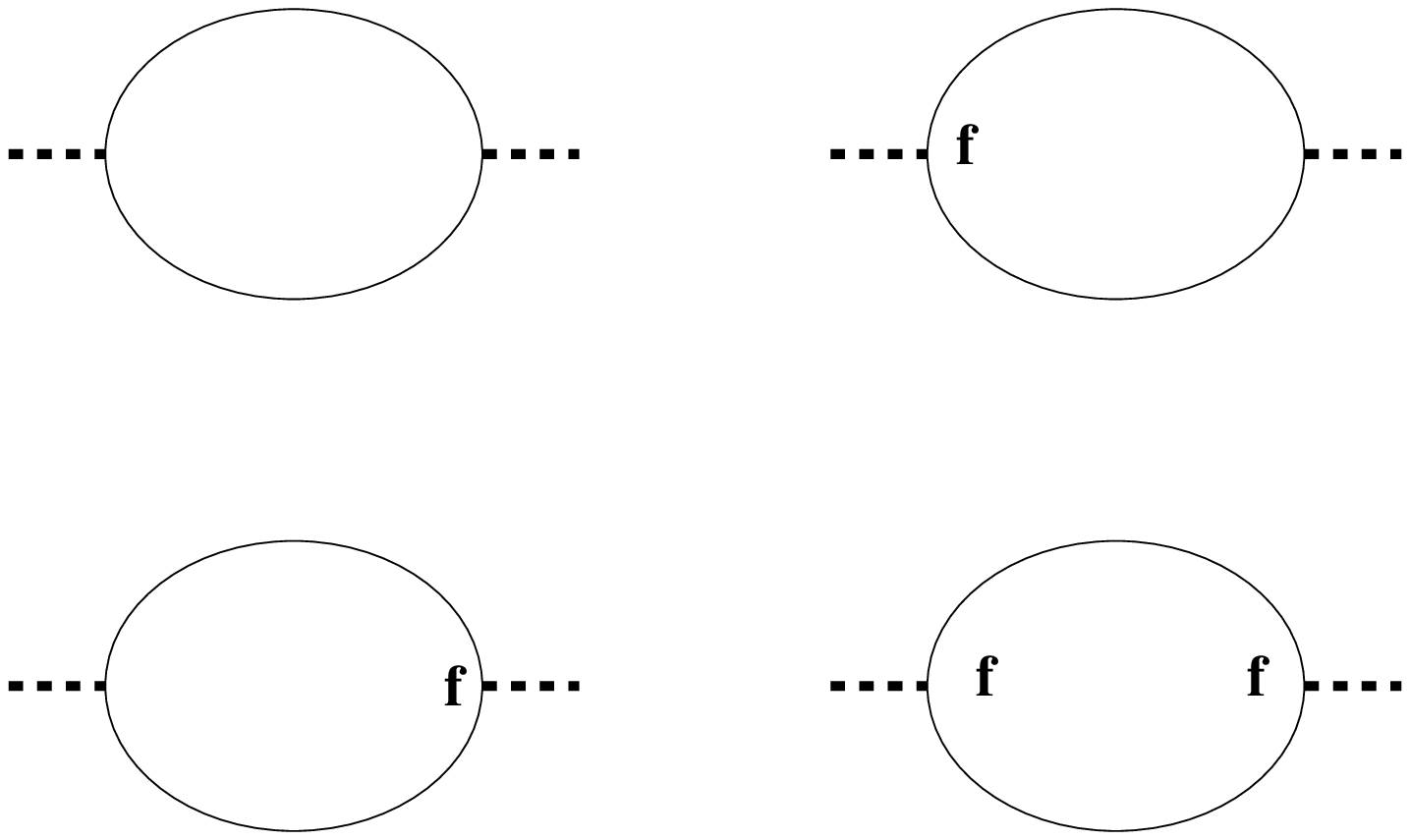}}
\caption{}
\label{loop1}
\end{figure}

\begin{figure}[t]
{\LARGE\bf $\pi_1 \times \partial_\mu$ }
\parbox[c]{4cm}{\psfig{file=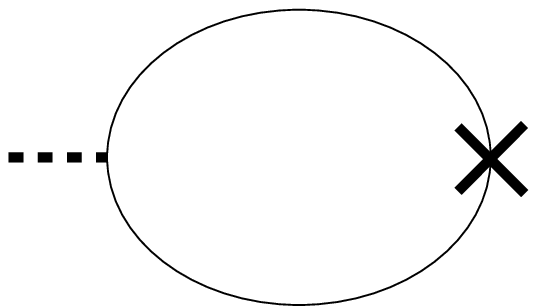}} 
{\LARGE\bf $ +  \pi_2 \times \partial_\mu  $}
\parbox[c]{4cm}{\psfig{file=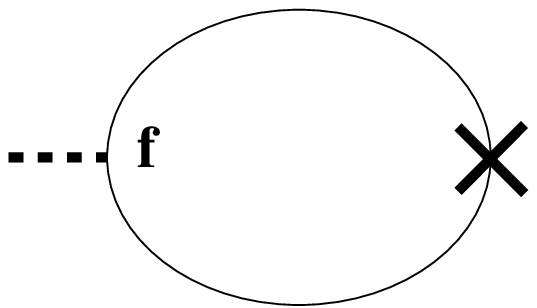}}
\caption{}
\label{pi-a}
\end{figure}

\begin{figure}[t]
  \centering
\psfig{file=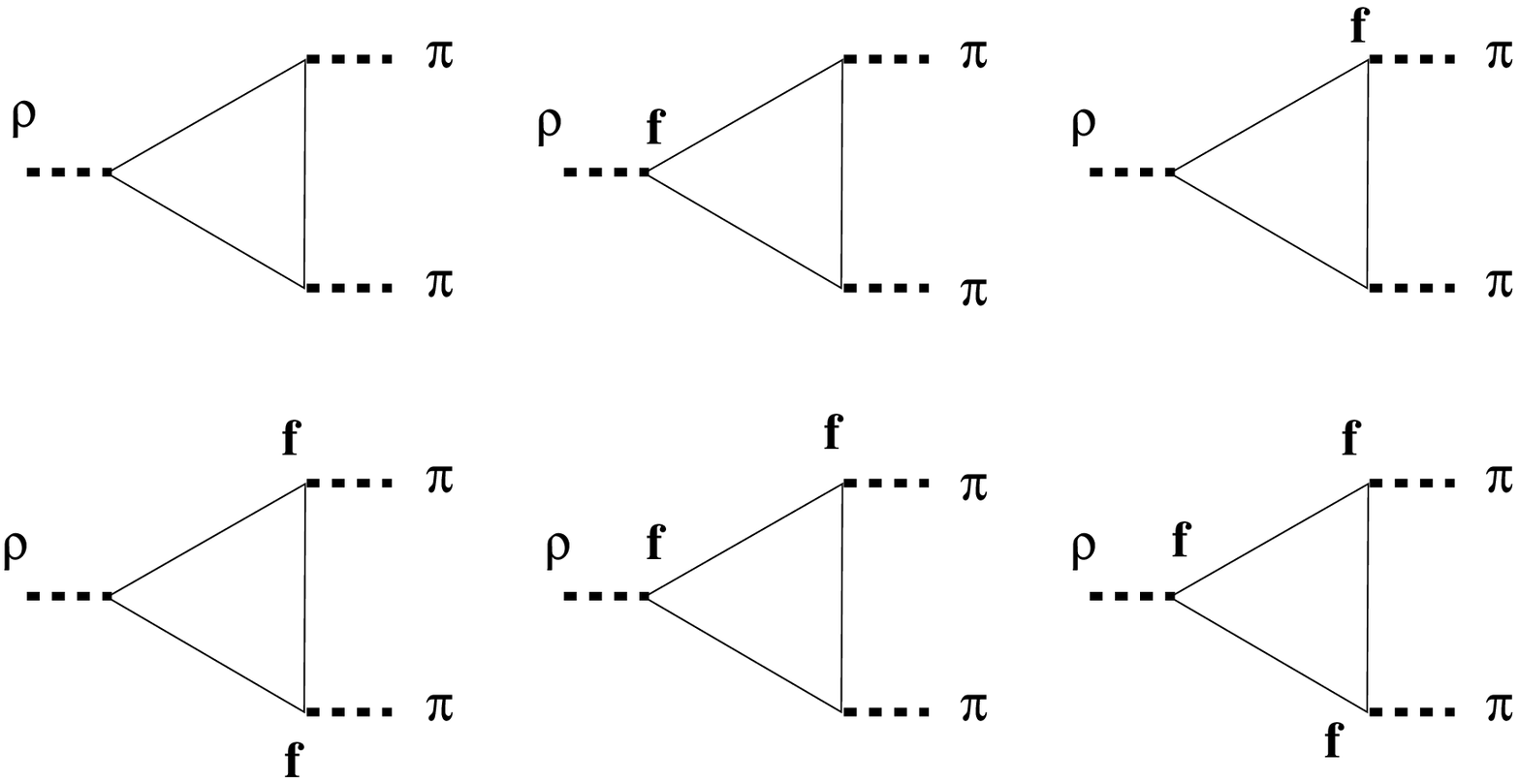}
\caption{}
  \label{fig:graph}
\end{figure}

\begin{figure}[t]
\begin{center}
\psfig{file=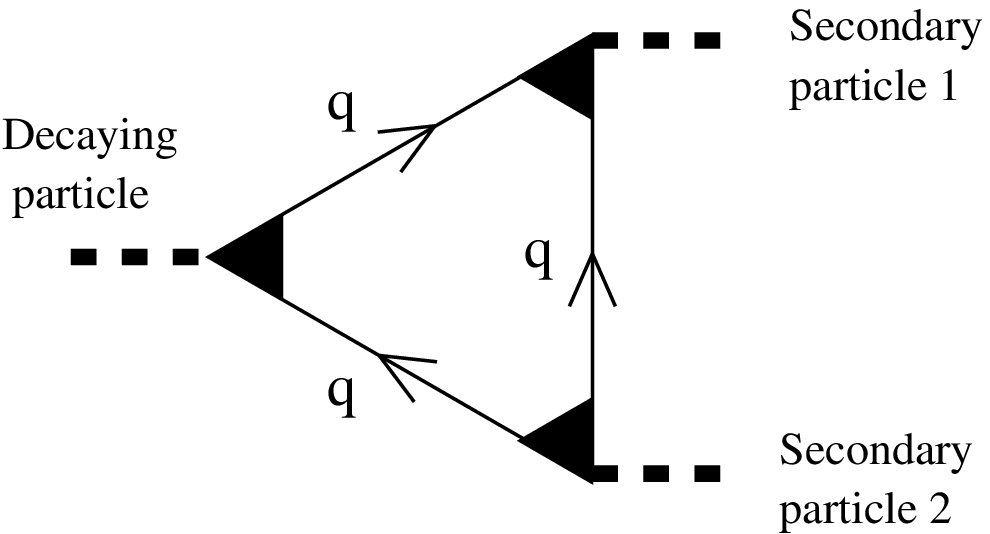}
\caption{}
\label{trigX}
\end{center}
\end{figure}

\begin{figure}[t]
\psfig{file=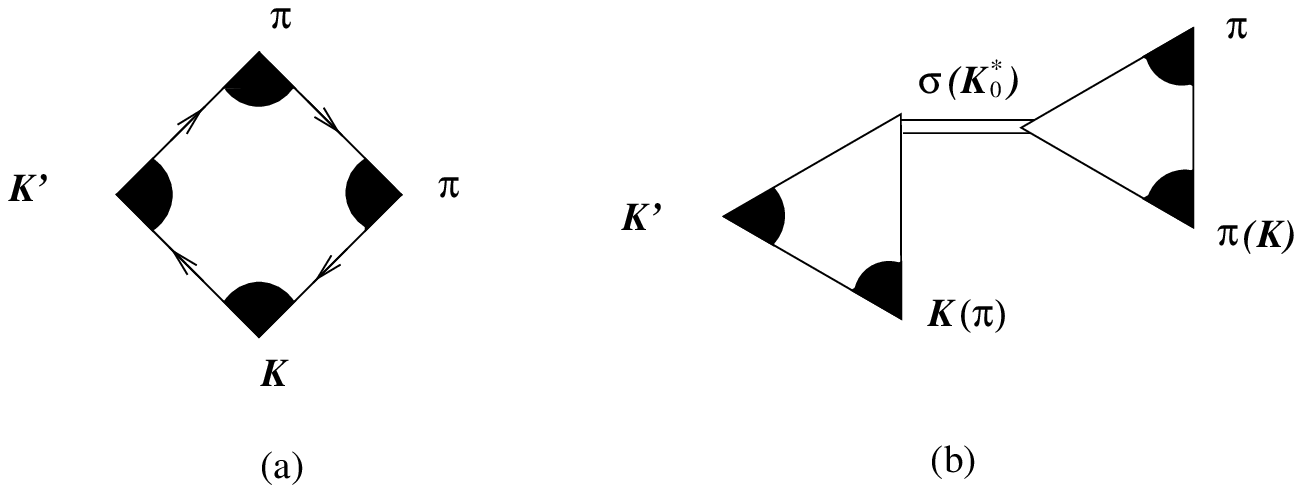}
\caption{ }
\label{KtoK2pi}
\end{figure}

\begin{figure}[t]
\begin{center}
\psfig{file=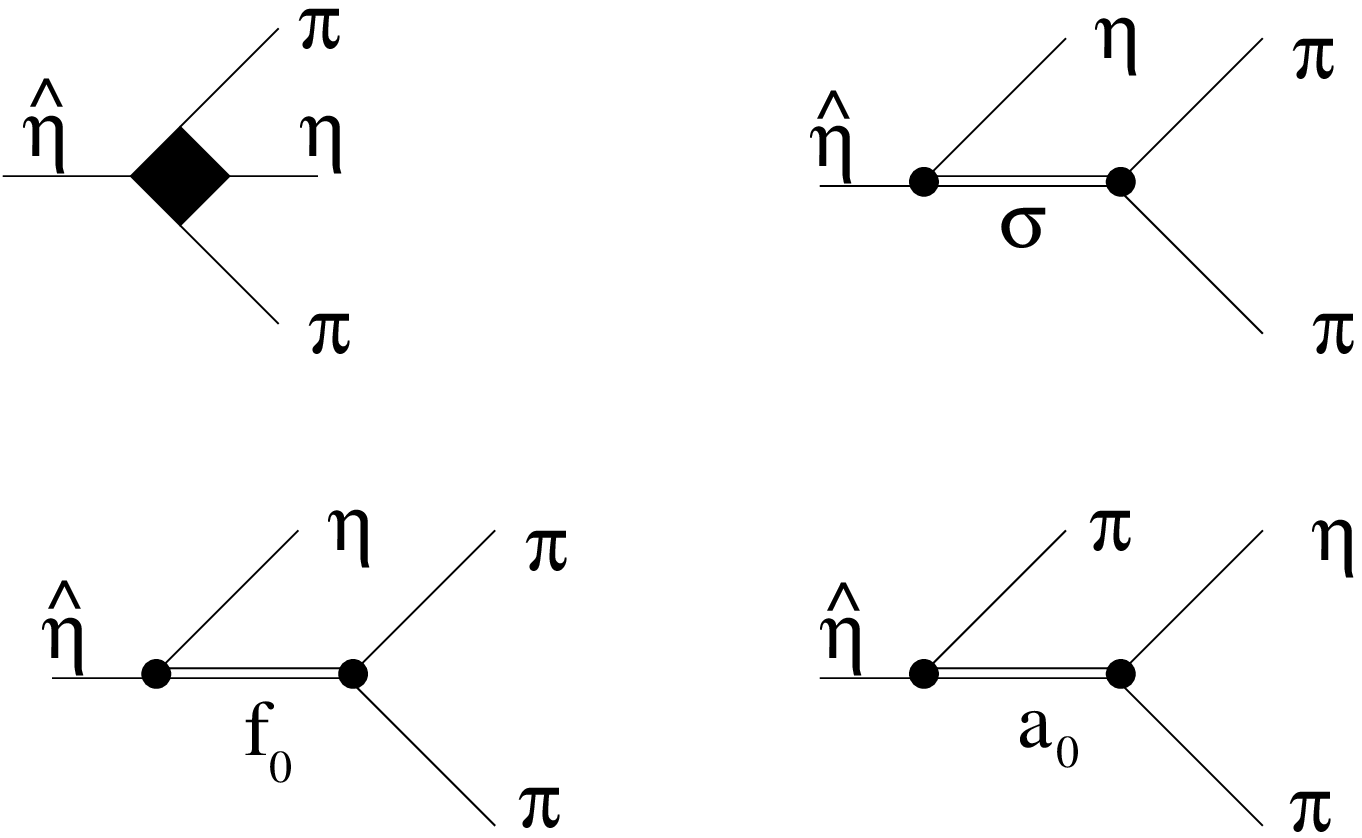}
\caption{  }
\label{polediag}
\end{center}
\end{figure}

\begin{enumerate}
\item The quark loop contribution to the quadratic form 
$K_{ij}(P)$, eq.(\ref{K_full}), of the effective action for 
$\pi_1$-- and $\pi_2$--fields. Solid lines denote the NJL quark
propagator. The $\pi_1$--field couples to  quarks through a local
vertex; the $\pi_2$--field, through the form factor, $f(k_\perp )$,
marked by letter {\bf f}.
\item The axial current of  $\pi_1$-- and $\pi_2$--fields,
Eq.(\ref{axial_current}), as it follows from the Noether theorem.  
The cross denotes a local axial current of quark fields to which
$\pi_1$-- and $\pi_2$--fields couple through quark
loops. The notation is the same as in Fig.~\ref{loop1}.
\item Triangle diagrams describing  decays of a $\rho$-meson. Each letter {\bf}
in a diagram indicates  the presence of a form factor at a vertex.
\item   Diagrams describing meson decays of the $1\to2$ type.
\item Diagrams describing the decay $K'\to K\pi\pi$.
\item  Diagrams describing the decay  $\hat\eta\to\eta\pi\pi$. The black box stands for
the sum of ``box'' diagrams represented by one-loop quark graphs with four meson vertices.
The rest of the  diagrams is a set of pole graphs with  $\sigma$, $f_0$, and $a_0$ scalar resonances.
 The diagram with $a_0$ is to be taken into account for two channels (due to the exchange of pions momenta).
 There are analogous contributions from radially excited resonances.
\end{enumerate}

\clearpage

%
%
\vfill
\eject
\clearpage
\vfill
\eject

\clearpage

\clearpage
\vfill
\eject


\clearpage

\clearpage

\end{document}